\documentclass[floats,floatfix,showpacs,amssymb,prd,twocolumn,superscriptaddress,nofootinbib,longbibliography,reprint,
]{revtex4-1}

\usepackage{soul}
\usepackage{dsfont}
\usepackage{graphicx}
\usepackage{dcolumn}
\usepackage{bm}
\usepackage{color}
\usepackage{epsfig}
\usepackage{hyperref}
\usepackage{natbib}
\usepackage{float}
\usepackage{footmisc}
\usepackage{aas_macros}
\usepackage{lipsum}
\usepackage{amsmath}
\usepackage{braket, xcolor}
\usepackage{hyperref}
\usepackage{comment}
\usepackage{MnSymbol}
\usepackage{enumitem}
\hypersetup{colorlinks, citecolor=blue, linkcolor=blue, urlcolor=blue}

\definecolor{lightred}{rgb}{1,0.5,0.5}
\definecolor{lightgreen}{rgb}{0.5,1,0.5}
\definecolor{lightblue}{rgb}{0.5,0.5,1}
\definecolor{lightcyan}{rgb}{0.5,0.75,0.75}
\definecolor{lightmagenta}{rgb}{0.75,0.5,0.75}
\definecolor{customgreen}{rgb}{0.494,1,0.502}

\newcommand{\be}{\begin{equation}}
\newcommand{\ee}{\end{equation}}
\newcommand{\beqa}{\begin{eqnarray}}
\newcommand{\eeqa}{\end{eqnarray}}
\newcommand{\dd}{{\rm d}}

\renewcommand{\vec}[1]{{\boldsymbol{#1}}}

\begin{document}

\title{LISA double white dwarf binaries as Galactic accelerometers}

\author{Reza Ebadi}
\email{ebadi@umd.edu}
\affiliation{Department of Physics, University of Maryland, College Park, Maryland 20742, USA}
\affiliation{Quantum Technology Center, University of Maryland, College Park, Maryland 20742, USA}

\author{Vladimir Strokov}
\email{vladimir.strokov@mail.wvu.edu}
\affiliation{The William H.~Miller III Department of Physics and Astronomy, The Johns Hopkins University, Baltimore, Maryland, 21218, USA}
\affiliation{Department of Physics and Astronomy, West Virginia University, Morgantown, WV 26506, USA}
\affiliation{Center for Gravitational Waves and Cosmology, West Virginia University, Chestnut Ridge Research Building, Morgantown, WV 26505, USA}

\author{Erwin H.~Tanin}
\email{ehtanin@stanford.edu}
\affiliation{Stanford Institute for Theoretical Physics, Stanford University, Stanford, California, 94305, USA}
\affiliation{The William H.~Miller III Department of Physics and Astronomy, The Johns Hopkins University, Baltimore, Maryland, 21218, USA}

\author{Emanuele Berti}
\email{berti@jhu.edu}
\affiliation{The William H.~Miller III Department of Physics and Astronomy, The Johns Hopkins University, Baltimore, Maryland, 21218, USA}

\author{Ronald~L.~Walsworth}
\email{walsworth@umd.edu}
\affiliation{Department of Physics, University of Maryland, College Park, Maryland 20742, USA}
\affiliation{Quantum Technology Center, University of Maryland, College Park, Maryland 20742, USA}
\affiliation{Department of Electrical and Computer Engineering, University of Maryland, College Park, Maryland 20742, USA}

\begin{abstract}
Galactic double white dwarf (DWD) binaries are among the guaranteed sources for the Laser Interferometer Space Antenna (LISA), an upcoming space-based gravitational wave (GW) detector. Most DWDs in the LISA band are far from merging and emit quasimonochromatic GWs. As these sources are distributed throughout the Milky Way, they experience different accelerations in the Galactic gravitational potential, and therefore each DWD exhibits an apparent GW frequency chirp due to differential acceleration between the source and LISA. We examine how Galactic acceleration influences parameter estimation for these sources; and investigate how LISA observations could provide insight into the distribution of matter in the Galaxy. 
\end{abstract}

\date{\today}

\maketitle

\section{Introduction \label{sec:intro}}
The Laser Interferometer Space Antenna (LISA)~\cite{2017arXiv170200786A,LISA:2022yao} is expected to detect the gravitational-wave (GW) signals of $\sim 10^4$ detached double white dwarfs (DWD) that reside primarily in the Milky Way galaxy~\cite{Lamberts:2019nyk,Korol:2017qcx,Korol:2021pun} and its satellites~\cite{Korol:2020lpq,Roebber:2020hso,Korol:2020hay,Keim:2022lzv, Rieck:2023pej}. Several tens of these DWDs have been identified through electromagnetic (EM) observations and are targets of particular significance for LISA, usually referred to as ``verification binaries''~\cite{Korol:2017qcx,2018MNRAS.480..302K,Littenberg:2024bso}.

One distinct advantage of GW observations is that they are unaffected by dust and stellar crowding, enabling probes of parts of the Galaxy that are challenging to access electromagnetically, such as those near or behind the Galactic Center~\cite{Wang:2020jsx,2020ApJ...904..113R, Wong:2019hsq,Sberna:2022qbn}. For this reason, several studies consider GWs from both the resolvable~\cite{Korol:2018wep,Georgousi:2022uyt} and unresolvable~\cite{Benacquista:2005tm,Breivik:2019oar} populations of DWDs as probes of Milky Way structures, complementary to EM observations. These studies propose using spectral and spatial features of the ensemble of Galactic GW signals from DWDs to infer the present \textit{spatial} distributions of Galactic DWDs, which are assumed to trace the underlying Milky Way structures (subject to inevitable biases from star formation rates, binary evolution, and observational selection). After the biases are modeled and removed, one can extract information about Milky Way structures, such as stellar mass, relative normalization, and morphology.

In addition to locating DWDs~\cite{Takahashi:2002ky}, GW measurements can also be used to determine the \textit{ relative motion} of resolved DWDs. While the motion of a DWD with constant velocity along the line of sight is completely degenerate with its intrinsic GW frequency, an \textit{apparent acceleration} along the line of sight can lead to a time-dependent Doppler shift (i.e., a frequency ``chirp'') in the emitted GWs that is, at least in principle, detectable~\cite{Yunes:2010sm,Bonvin:2016qxr,Inayoshi:2017hgw,Meiron:2016ipr,Robson:2018svj,Tamanini:2019usx,Randall:2018lnh,Wong:2019hsq,Strokov:2021mkv,Vijaykumar:2023tjg}.
We stress that, here and throughout the paper, what we refer to as the ``apparent acceleration'' along the line of sight $\vec{\hat{n}}$ of a DWD at a distance $D$ from the Sun is the sum of the its line-of-sight acceleration $\vec{a}\cdot\vec{\hat{n}}$ and the perspective acceleration $\mu^2D$ due to its proper motion $\mu$ on the sky (also known as the ``Shklovskii effect''~\cite{1970SvA....13..562S}).

Provided that one can measure both the location and the apparent acceleration of a DWD, there is the tantalizing prospect of effectively using a population of DWDs as ``test mass'' accelerometers to probe Galactic structures. The motions of DWDs inferred from GW measurements encode different and, in principle, more direct information about the underlying Galactic gravitational potential compared to what can be inferred from their spatial distribution alone. There is, however, an important challenge to the practical application of this technique: GW phase changes due to the apparent acceleration of a DWD can be confused with its intrinsic frequency chirp due to GW emission. That said, additional information from EM observations of the same sources (before, during, and after GW measurements) may be used to break this degeneracy.

Motivated by the prospect of Galactic accelerometry via LISA measurements of GWs from DWDs--and by the anticipation of finding the EM counterparts to LISA sources--in this paper we address the following questions:

\begin{itemize}
  \item[(1)] Under what circumstances, and to what degree, can we recover the apparent acceleration of individual DWDs from their gravitational waveforms, as observed by LISA?
  \item[(2)] How well can we characterize large-scale Milky Way mass profiles (distributions) by including in the analysis the apparent acceleration of the large number of DWDs that LISA will resolve?
\end{itemize}

In the process, we also shed light on how the inclusion of apparent acceleration in the GW template affects the inferred DWD binary parameters.

The rest of the paper is organized as follows. In Sec.~\ref{sec:accel_map} we provide a model map of the Galactic acceleration and describe the synthetic DWD population used for subsequent estimates. In Sec.~\ref{sec:fisher} we discuss the measurability of the center-of-mass acceleration of a DWD by carrying out a Fisher matrix analysis with GW observations alone. Then we consider the case in which some parameters of the source are assumed to be known from EM observations. In Sec.~\ref{sec:accelerometry} we consider the prospect of using LISA for detection of Galactic acceleration, and of leveraging the large number of observable DWDs to improve measurement uncertainties. In Sec.~\ref{sec:conclusion} we present our conclusions.
To improve readability, some technical material is relegated to the Appendices.
In Appendix~\ref{app:shklovskii} we briefly review the notion of perspective acceleration due to the proper motion of a source on the sky (the ``Shklovskii effect''). In Appendix~\ref{app:fisher} we perform a comparison of frequency domain (FD) and time domain (TD) results, showing that they are in good agreement.

Table\,\ref{tab:notation} provides a list of frequently used symbols used in this paper and of their definitions.
In what follows we use geometrical units $G=c=1$, where $G$ is the gravitational constant and $c$ is the speed of light.

\begin{table}[hbtp!]
\small
\caption{List of frequently used symbols. Variations thereof and symbols used in passing are defined in the text.}\label{tab:notation}%
\begin{center}
\begin{tabular}{l|c|l}\hline\hline
\begin{tabular}{@{}c@{}}Quantity\\ \end{tabular} & \begin{tabular}{@{}c@{}}Symbol \\ \end{tabular} & \begin{tabular}{@{}c@{}}Definition \\ \end{tabular} \\ \hline\hline
{} & {} & {} \\
Galactocentric radius & $r$ & $\sqrt{x^2+y^2+z^2}$ \\
\begin{tabular}{@{}l@{}} Galactocentric radius projected \\ onto the Galactic plane  \end{tabular}  & $R$ & $\sqrt{x^2+y^2}$\\
Source distance from Sun & $D$ & {} \\
Galactic gravitational potential & $\Phi$ & {} \\
Galactic potential normalization & $\mathcal{N}$ & $\Phi=\mathcal{N}\Phi_\mathrm{model}$ \\
Line-of-sight unit vector & $\vec{\hat{n}}$ &  {} \\
Total source velocity & $v$ & {} \\
Source velocity in the sky plane & $v_\perp$ &  
\\
Local acceleration & $\vec{a}_\odot$ & {} \\
Galactic line-of-sight acceleration & $a_\parallel$ & $(-\vec{\nabla}\Phi-\vec{a}_\odot)\cdot\vec{\hat{n}}$ \\
Perspective acceleration & -- & $v_\perp^2/D$ \\
Apparent acceleration & $a$ & $a_\parallel+v_\perp^2/D$ \\
Observer frame time & $t$ & {}\\
Source frame time & $\tau$ & {}\\
Coalescence time & $\tau_\mathrm{c}$ & {} \\
Observer frequency & $f$ & {}\\
Source frequency & $f_\mathrm{s}$ & {}\\
GW phase & $\psi$ & {} \\
Coalescence phase & $\psi_\mathrm{c}$ & {} \\
DWD chirp mass & $\mathcal{M}$ & {} \\
Time domain waveform & $h(t)$ & {} \\
Frequency domain waveform & $h_F$ & {} \\
Fourier frequency & $F$ & {} \\
LISA observation time & $T_\mathrm{obs}$ & {} \\
Error of $\theta_i$ in parameter inference & $\sigma_{\theta_i}$ & {} \\
Fractional error & $\varepsilon_{\theta_i}$ & $\sigma_{\theta_i}/\theta_i$ \\
{} & {} & {} \\

\hline
\end{tabular}
\end{center}
\end{table}

\section{Galactic acceleration and the DWD population \label{sec:accel_map}}

In this section we present a simplified model of the Milky Way gravitational potential, from which we produce a model map of the Galactic acceleration contributing to the frequency chirp of GW sources. We also describe a synthetic population of Galactic DWDs that is used for Fisher calculations presented in subsequent sections.

\subsection{Galactic acceleration}\label{sec:gal_accel_A}

\begin{figure*}[t]
    {\includegraphics[width=1\textwidth]{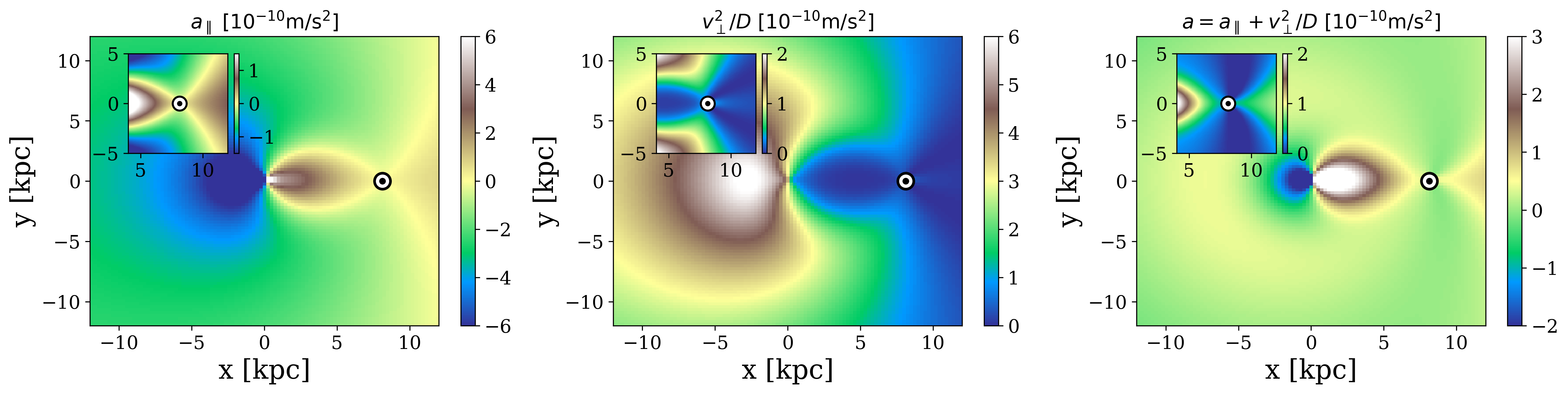}}
    \caption{Model maps of Galactic acceleration. Left: Galactic line-of-sight acceleration contribution $\Delta\vec{a}\cdot\hat{n}$. Center: Shklovskii (perspective) contribution using only $v_\mathrm{circ.}$. Right: Sum of the two contributions. The contours of these quantities are shown as functions of $x$ and $y$, the Galactocentric coordinates on the Galactic plane, aligned such that the location of the Sun (labeled by $\odot$) is along the $x$ axis. Insets in each panel show acceleration maps centered on the Sun, with enhanced contrast.}%
    \label{fig:accel}%
\end{figure*}

We assume a model for the coarse-grained gravitational potential of the Milky Way, $\Phi_{\rm model}$, which comprises a stellar disk, a bulge, and a dark matter (DM) halo: 
\begin{align}
    \Phi_{\rm model}= \Phi_{\rm disk} + \Phi_{\rm bulge} + \Phi_{\rm DM} \label{eq:Phimodel}\,,
\end{align}
where the three components $\Phi_{\rm disk}$, $\Phi_{\rm bulge}$, and $\Phi_{\rm DM}$ are respectively modeled as a Miyamoto--Nagai stellar disk~\cite{1975PASJ...27..533M}, a Hernquist sphere bulge~\cite{1990ApJ...356..359H}, and a Navarro--Frenk--White DM halo~\cite{1996ApJ...462..563N}:
\begin{subequations}\label{eq:Phi_comps}
\begin{equation}
  \Phi_{\rm disk}=-\frac{M_{\rm disk}}{\sqrt{r^2+(c_1+\sqrt{z^2+c_2^2})^2}}\,,\label{eq:Phidisk} 
\end{equation}    
\begin{equation}
  \Phi_{\rm bulge}=-\frac{M_{\rm bulge}}{r+c_3}\,,\label{eq:Phibulge}
\end{equation}
\begin{equation}
  \Phi_{\rm DM}=-\frac{4\pi \rho_0 r_s^3}{r}\ln\left(1+\frac{r}{r_s}\right)\,.\label{eq:PhiDM}
\end{equation}
\end{subequations}
Here $(x,y,z)$ are the Galactocentric coordinates, and $r=\sqrt{x^2+y^2+z^2}$. The free parameters are assumed to be $M_{\rm disk}=7\times 10^{10}M_{\odot}$\,, $c_1=3\mathrm{\,kpc}$, $c_2=0.28\,\mathrm{kpc}$\,, $M_{\rm bulge}=5\times 10^9M_{\odot}$\,, $c_3=0.6\,\mathrm{kpc}$\,, $\rho_0=9\times 10^{-3}\,M_{\odot}\;\mathrm{pc}^{-3}$\,, and $r_s=16\,\mathrm{kpc}$~\cite{2015ApJS..216...29B,Bovy:2013raa}.

Given this model of the Milky Way gravitational potential, the Galactocentric acceleration of a source is simply $-\vec{\nabla}\Phi_{\rm model}$. In the observer frame, a time-dependent Doppler modulation of the GW phase results from the line-of-sight component of the \textit{apparent} acceleration of the source. We assume for simplicity that the motion of the observer follows that of the Sun, with position, velocity, and acceleration relative to the Galactic center given by $\vec{R}_{\odot}=8.12\,\mathrm{kpc}\,\vec{\hat{x}}$~\cite{2018A&A...615L..15G}, $\vec{v}_\odot= 223\,\mathrm{km/s}\,\vec{\hat{y}}$, and $\vec{a}_\odot=\left.\vec{\nabla}\Phi\right|_{\vec{R}_\odot}=-1.9\times10^{-10}\,\mathrm{m/s^2}\,\vec{\hat{x}}$, respectively. (Small corrections to account for the motion of LISA relative to the Sun can be made in future work.) Here we assume that the Sun follows a circular orbit, with velocity given by the Galactic rotation curve: 
\begin{align}\label{eq:v_circ_from_Phi}
    v^2=v_{\rm circ}^2=R\frac{\partial \Phi}{\partial R}\,,
\end{align}
with $R=\sqrt{x^2+y^2}$ being the Galactocentric radius projected onto the Galactic plane.

The apparent acceleration of a Galactic source has two contributions: the line-of-sight projection of the Galactic acceleration w.r.t. the observer, $(-\vec{\nabla}\Phi_{\rm model}-\vec{a}_\odot)\cdot\vec{\hat{n}}\equiv(\dot{\vec{v}}-\dot{\vec{v}}_\odot)\cdot\vec{\hat{n}}$, and the perspective acceleration $a_\mathrm{Sh} = \mu^2 D$ due to the source's proper motion $\mu=v_\perp/D$ on the sky (otherwise known as the Shklovskii effect~\cite{1970SvA....13..562S}); here, $\vec{\hat{n}}$ is the unit vector along the line-of-sight, $D$ is the distance to the source, and $v_\perp$ is the component of the source's velocity w.r.t. to the observer perpendicular to $\vec{\hat{n}}$. In order to estimate $v_\perp$, and thus the contribution of the perspective acceleration, we distribute DWDs in the stellar disk and bulge following Refs.~\cite{Korol:2017qcx,Korol:2021pun}; and assume that the DWDs follow near-circular orbits in accordance with Eq.~\eqref{eq:v_circ_from_Phi}. Hence, the apparent acceleration reads
\begin{align}
    a=(-\vec{\nabla}\Phi_{\rm model}-\vec{a}_\odot)\cdot\vec{\hat{n}} +\frac{v_{\perp}^2}{D}\,,
\end{align} 
where $v_\perp$ is the component of $\vec{v}_{\rm circ}-\vec{v_{\odot}}$ perpendicular to $\vec{\hat{n}}$, and the vector $\vec{v}_{\rm circ}$ is tangential to the orbit of the source.

\begin{figure*}[htpb!]%
    \includegraphics[width=0.55\textwidth]{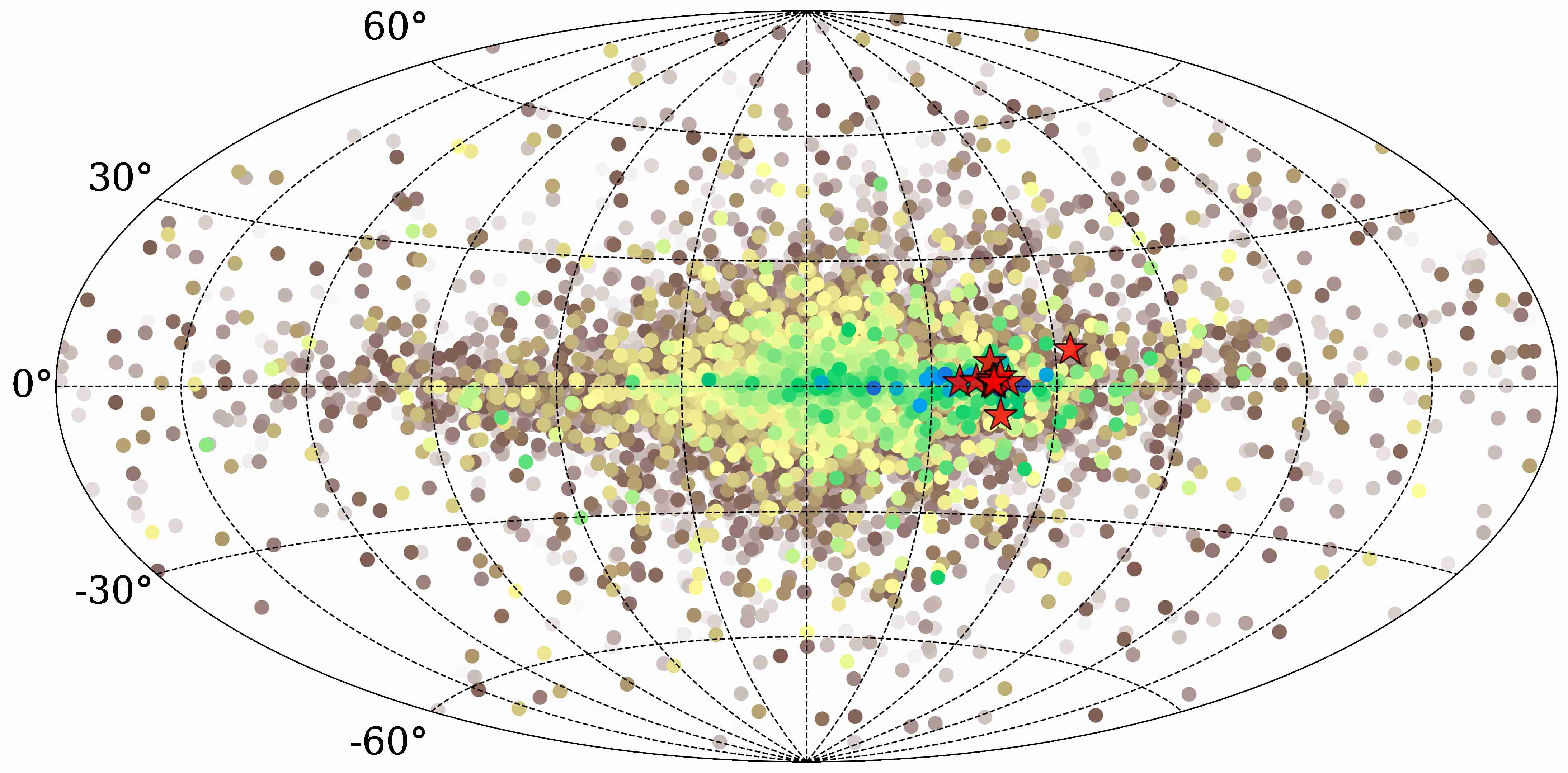}%
    \includegraphics[width=0.4\textwidth]{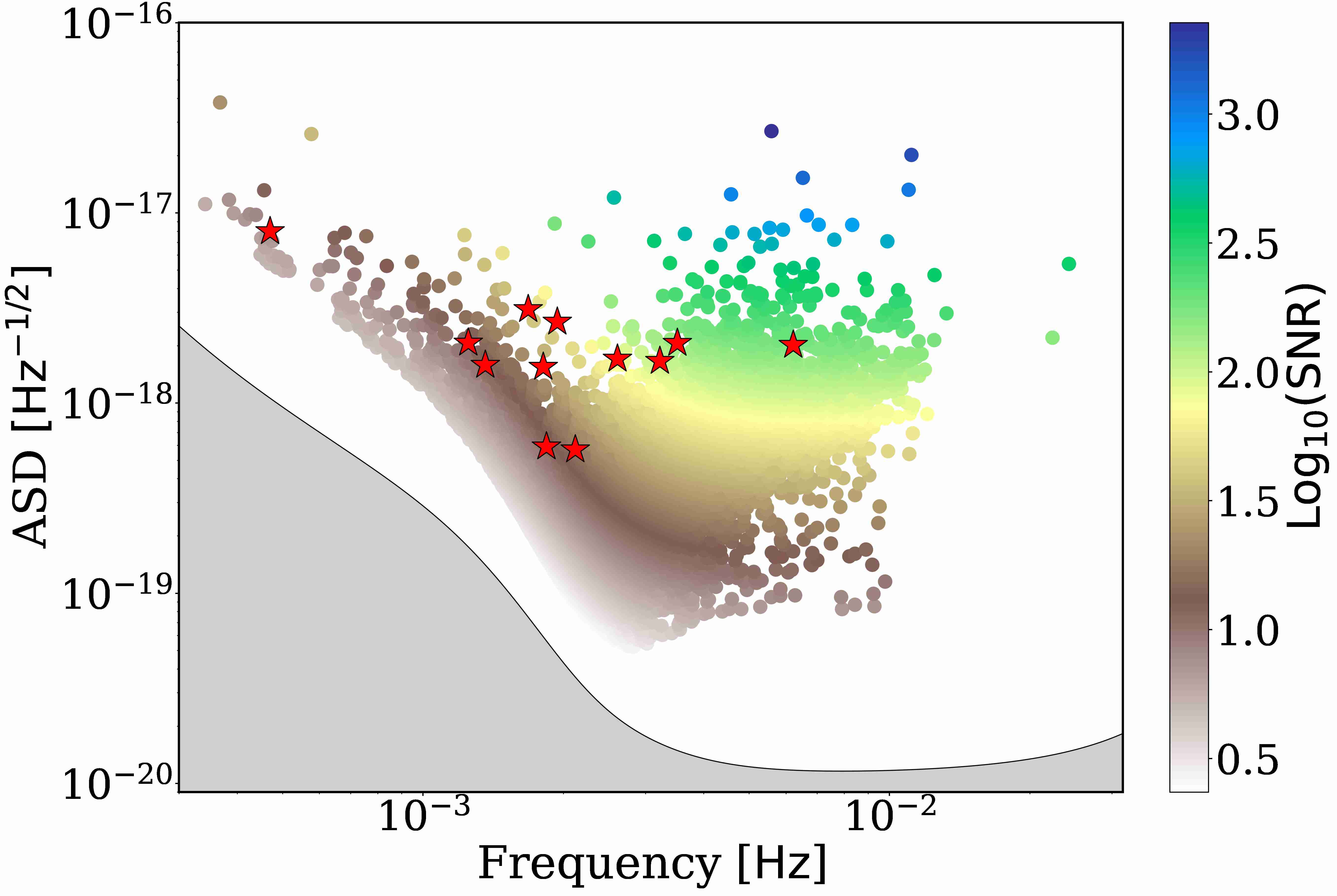}%
    \caption{Synthetic population of Galactic DWDs observable in the 10\,yr LISA mission. {\it Left:} Aitoff projection of the population in the Galactic longitude and latitude map. {\it Right:} The population shown in the amplitude spectral density (ASD) vs. frequency plane, together with the LISA noise curve (in gray), which includes the DWD confusion noise. DWD sources are simulated following Ref.~\cite{Thiele:2021yyb} (the total number of sources is 16264); verification binaries (labeled with red stars) are from Ref.~\cite{2018MNRAS.480..302K}. For both panels, DWD symbol color indicates LISA SNR for the associated GW.}%
    \label{fig:sky_lisa}%
\end{figure*}

Figure\,\ref{fig:accel} shows maps of the line-of-sight acceleration $a_\parallel = (-\vec{\nabla}\Phi_{\rm model}-\vec{a}_\odot)\cdot\vec{\hat{n}}$, perspective acceleration $v_\perp^2/D$, and their sum $a$ (the apparent acceleration) following from the model of the Milky Way potential given by Eqs.~(\ref{eq:Phimodel}) and (\ref{eq:Phi_comps}). The 
$x\leftrightarrow -x$ asymmetry of the maps is due to the off-center position of the Solar System, where the observer (e.g., LISA) resides.
The line-of-sight acceleration is negative on the far side of the Galactic center ($x<0$) and can be either positive or negative on the observer's side ($x>0$). The magnitude of the acceleration generally grows towards the Galactic Center and decreases near the Sun. The perspective acceleration is always positive: it is strongest near the Galactic Center on the far side, where the $y$-component of the circular velocity is opposite to that of the Sun; and weakest in the Sun's vicinity, where the relative velocity with respect to the Sun almost vanishes. Depending on location, the two contributions to the total acceleration may amplify or cancel each other. Most importantly, the typical apparent acceleration is of the order of $\sim 10^{-10}\text{ m}/\text{s}^2$.

In general, stellar populations also undergo random peculiar motions $\vec{v}_p$ relative to the local $\vec{v}_{\rm circ}$, with the typical components of $\vec{v}_p$ given by their respective velocity dispersions. For instance, the Sun has an oblique peculiar velocity of magnitude $v_p=10.5\pm 1.5\text{ km/s}$ relative to the circular rotation velocity at the Sun $v_{\rm circ, \odot}=238\pm 15\text{ km/s}$ (the local standard of rest)~\cite{2016ARA&A..54..529B}. Stars in the disk are subject to a multitude of scattering mechanisms that increase their velocity dispersions over time~\cite{2020AJ....160...43A}, and consequently
the random stellar motions of the current stellar populations depend strongly on their age, metallicity, and locations of the population in question~\cite{2021MNRAS.506.1761S}. Among the assortment of stars, the kinematics of white dwarfs is relatively less understood, since their characteristically strong surface gravity and pressure lead to line broadening that makes the extraction of their radial velocities challenging~\cite{2022A&A...658A..22R}. The reported values of the RMS sum of the velocity dispersions in different directions of both isolated white dwarfs and DWDs in the disk typically do not exceed $60\text{ km/s}$~\cite{2012MNRAS.426..427W, 2015ApJ...812..167G, 2017MNRAS.469.2102A, 2022A&A...658A..22R}, which amounts to $\lesssim 5\%$ corrections to the perspective acceleration when added in quadrature to the $\sim 250\text{ km/s}$ circular rotation velocity at $R\gtrsim 5\text{ kpc}$~\cite{2016ARA&A..54..529B}. Since the peculiar motion of DWDs appears to have little effect on their perspective accelerations, we neglect it in our analysis. The peculiar motion of stars may also slightly modify the mean azimuthal velocity of the population at a given location -- an effect known as asymmetric drift~\cite{2016ARA&A..54..529B}, which we also neglect. 

Also, while we adopt a simplified equilibrium and axisymmetric Galactic model in our analysis, it is known that the Milky Way exhibits non-equilibrium and non-axisymmetric properties, such as stellar streams, and bar and spiral structures of the stellar disk~\cite{2013NewAR..57...29B}. For example, in the $N$-body model GALAKOS of a Milky Way-like galaxy~\cite{2020ApJ...890..117D}, the contribution of a bar and spiral arms to the number density is at most comparable with the axisymmetric part~(see Figure~6 of Ref.\cite{2021MNRAS.500.4958W}), and therefore it does not affect the order of magnitude obtained assuming axial symmetry. Once the main effect is detected (i.e., acceleration contribution to the frequency chirp of DWD GW signals), these details can be taken into account in future work through more realistic models of the Milky Way.

\subsection{Synthetic DWD population\label{subsec:DWDpopulation}}

Simulations of the Galactic binary population in the LISA band have improved in recent years~\cite{2001A&A...365..491N,2013MNRAS.429.1602Y,Korol:2017qcx,2019MNRAS.490.5888L,Thiele:2021yyb,2022ApJ...937..118W}. These simulations typically predict $\mathcal{O}(10^4)$ resolvable DWDs, making them the most abundant LISA sources. Additionally, dozens of black hole and neutron star binary systems are expected to be detected by LISA~\cite{2022ApJ...937..118W}. In this work we consider a synthetic population of DWD GW sources following Thiele \textit{et al.}~\cite{Thiele:2021yyb}, where the authors use simulations of Milky Way-like galaxies and stellar evolution codes to generate a present-day Galactic binary population. We adopt their model~FZ~\cite{Thiele:2021yyb}, which relies on a metallicity-dependent initial binary fraction, as suggested by electromagnetic observations in the solar neighborhood. Due to tidal effects and mass transfer at play in the progenitor binaries, all DWDs in the population are circular. This gives a typical GW emission timescale of $\gtrsim 10^4$~yr. 

Figure\,\ref{fig:sky_lisa} illustrates the characteristics of the synthetic DWD population. We choose the resolvable sources for a 10 yr LISA mission by requiring a position-, orientation- and polarization-angle averaged $\mbox{SNR}>7$, which results in about $16,000$ resolvable sources. The subset of the DWD population usually referred to as ``verification binaries'' consists of sources known from prior EM observations, and guaranteed to generate detectable GW signals in LISA. The current collection of verification binaries includes $\sim10$ detached double white dwarfs~\cite{2018MNRAS.480..302K, Finch:2022prg,Kupfer:2023nqx} expected to have $\text{SNR}>7$, and marked by red stars in Figure~\ref{fig:sky_lisa}. 

\section{Fisher analysis \label{sec:fisher}}

The ubiquity of DWDs suggest they are well suited to serving as Galactic accelerometers, enabling mapping of the Galactic gravitational potential. However, as we first motivate with a ``back-of-the-envelope'' argument (Section~\ref{subsec:QMS}), it is challenging to determine the center-of-mass acceleration of a DWD via GW measurements. In Section~\ref{subsec:FD} we confirm the back-of-the-envelope estimates with a Fisher matrix analysis for a few representative cases. 

\subsection{DWDs as accelerometers\label{subsec:QMS}}

If a DWD is moving with an apparent acceleration~$a\equiv\dot{v}_{||}$, the time dependence of the GW frequency $f(t)$ registered by LISA (i.e., at the ``observer'') will be different from the intrinsic chirping $f_{\rm s}(\tau)$ of the GW signal emitted by the DWD (i.e., at the ``source''). We will assume that the observer's clock and the clock at the source are synchronized at the beginning of the observation -- i.e., a GW signal that the observer starts receiving at the initial time~$t=0$ was emitted at $\tau=0$ by the source's clock. In this work we only consider effects linear in the (assumed small) DWD acceleration~$a$. For simplicity we also assume the DWD orbits to be circular, and their inspiral to be driven purely by gravitational radiation. 

The GW frequency drift exprienced by DWDs is slow, as evidenced by their long coalescence times:
\beqa
\tau_{\rm c}&=&\frac{5\mathcal{M}}{256}(\pi \mathcal{M}f_{\rm s0})^{-8/3}\nonumber\\
&\approx& 1.3\times 10^{6}\;\mbox{yr}\left(\frac{\mathcal{M}}{0.44M_\odot}\right)^{-5/3}\left(\frac{f_{\rm s0}}{2\text{ mHz}}\right)^{-8/3},
\label{eq:coalesc-time}
\eeqa
where $f_{\rm s0}=\left.f_{\rm s}\right|_{\tau=0}$ is the initial frequency of the source, and $\mathcal{M}=(m_1m_2)^{3/5}(m_1+m_2)^{-1/5}$ is the chirp mass, a function of the individual masses $m_1$ and $m_2$ of the binary components (e.g., $\mathcal{M}=0.44M_\odot$ for $m_1 = m_2 = 0.5M_\odot$). From the (long) coalescence time and the nominal LISA observation time $T_{\rm obs}=10$~yr~\cite{2017arXiv170200786A,2021arXiv210709665S,LISA:2022yao}
we can define a small parameter $\tau/\tau_{\rm c} \leq T_{\rm obs}/\tau_{\rm c} \sim 10^{-5}$, which gives rise to the expansion of the GW frequency at the source:
\be
f_{\rm s}(\tau)=f_{\rm s0} + \dot{f}_{\rm s0}\tau + \ddot{f}_{\rm s0}\tau^2/2+\ldots\,.
\ee 
Indeed, if we estimate the derivatives $\dot{f}_{\rm s0}\sim f_{\rm s0}/\tau_{\rm c}$, $\ddot{f}_{\rm s0}\sim f_{\rm s0}/\tau_{\rm c}^2,\ldots$\,, the respective terms are $\dot{f}_{\rm s0}\tau\lesssim T_{\rm obs}/\tau_{\rm c}$, $\ddot{f}_{\rm s0}\tau^2\lesssim (T_{\rm obs}/\tau_{\rm c})^2$, etc. More precisely, at the lowest post-Newtonian order~\cite{Peters:1963ux,Peters:1964zz},
\beqa
f_{\rm s}(\tau) &=& f_{\rm s0}\left(1-\frac{\tau}{\tau_{\rm c}}\right)^{-3/8} \nonumber \\
&=& f_{\rm s0} + \frac 38(f_{\rm s0}/\tau_{\rm c})\,\tau + \frac{33}{128}(f_{\rm s0}/\tau_{\rm c}^2)\,\tau^2 + \ldots \\
&\Rightarrow& {} \quad \dot{f}_{\rm s0} = \frac 38\frac{f_{\rm s0}}{\tau_{\rm c}}\,, \quad \ddot{f}_{\rm s0} = \frac{33}{64}\frac{f_{\rm s0}}{\tau_{\rm c}^2}\,, \quad \ldots\,. \label{eq:exact-fdots}
\eeqa 

Apparent DWD acceleration causes the observed derivatives $\dot{f}_0$, $\ddot{f}_0,\ldots$\,, to differ from their counterparts at the source. In analogy with the well-known degeneracy between $f_{\rm s0}$ and a constant relative velocity~$v_{||}$ (and the chirp mass), the first derivative $\dot{f}_{\rm s0}$ is fully degenerate with the apparent acceleration, as long as the motion is nonrelativistic. This can be illustrated by a source that is receding just fast enough to compensate for the GW frequency drift: the linear growth of the frequency due to the chirp $\dot{f}_{\rm s0}$ is balanced by the Doppler shift, which is also linear in case of a constant acceleration. Therefore, in order to break this degeneracy and determine the apparent acceleration from the GW signal alone, it is necessary to measure at least the second derivative $\ddot{f}_0$.      

We now consider the observed frequency $f(t)=f_0 + \dot{f}_0 t + \ddot{f}_0t^2/2+\ldots$ and estimate how realistic it is to measure, more generally, the $k$th derivative~$f^{(k)}_0$. We do so by calculating the respective dephasing, i.e., the contribution of a given derivative term to the GW phase
\be
\psi(t) = 2\pi\int\limits_0^{t}{f(t')\,\dd{t'}} = 2\pi \left(f_0 t + \frac{1}{2!}\dot{f}_0 t^2 + \frac{1}{3!}\ddot{f}_0 t^3 + \ldots\right)\,. \label{eq:generalTDphase}
\ee
Of course, the GW signal must be loud enough to be detected, but this is always the case for the population of resolved DWDs with $\mbox{SNR}>7$ (see Section~\ref{subsec:DWDpopulation}). The  contribution to the dephasing accumulated by the term proportional to $f_0^{(k)}$ during the observation time~$T_{\rm obs}$ can be estimated from:
\be
\label{eq:psik_deriv}
\delta\psi_k\equiv  \frac{2\pi}{(k+1)!}f_0^{(k)}T_{\rm obs}^{k+1} = \frac{2\pi\,\delta f_k\,T_{\rm obs}}{k+1}\gtrsim 2\pi\,,  
\ee
where we introduce the respective frequency drift $\delta f_k=f^{(k)}T_{\rm obs}^k/k!$\,, which is likely to be detected if it exceeds the typical size of a frequency bin, $\delta f_k\gtrsim 1/T_{\rm obs}$. Once again, estimating the derivatives as $f^{(k)}\sim f_0/\tau_{\rm c}^k$, we recast the above formula in terms of the zeroth-order contribution:
\be
\label{eq:psik_tc}
\delta\psi_k \sim \frac{2\pi f_0 T_{\rm obs}}{(k+1)!}\left(\frac{T_{\rm obs}}{\tau_{\rm c}}\right)^k \sim 10^{6-5k}\left(\frac{f_0 T_{\rm obs}}{10^6}\right)\left(\frac{T_{\rm obs}/\tau_{\rm c}}{10^{-5}}\right)^k\,,
\ee
where the normalization values are typical of a $0.5M_\odot+0.5M_\odot$ DWD that emits at $2\,\mbox{mHz}$ (see Eq.~(\ref{eq:coalesc-time})). For such a binary, the chirp $\dot{f}_0$ is likely measurable ($\delta\psi_1\gtrsim 1$), whereas the second derivative~$\ddot{f}_0$ is not ($\delta\psi_2\ll 1$).

\begin{figure}[t]
    {\includegraphics[width=\columnwidth]{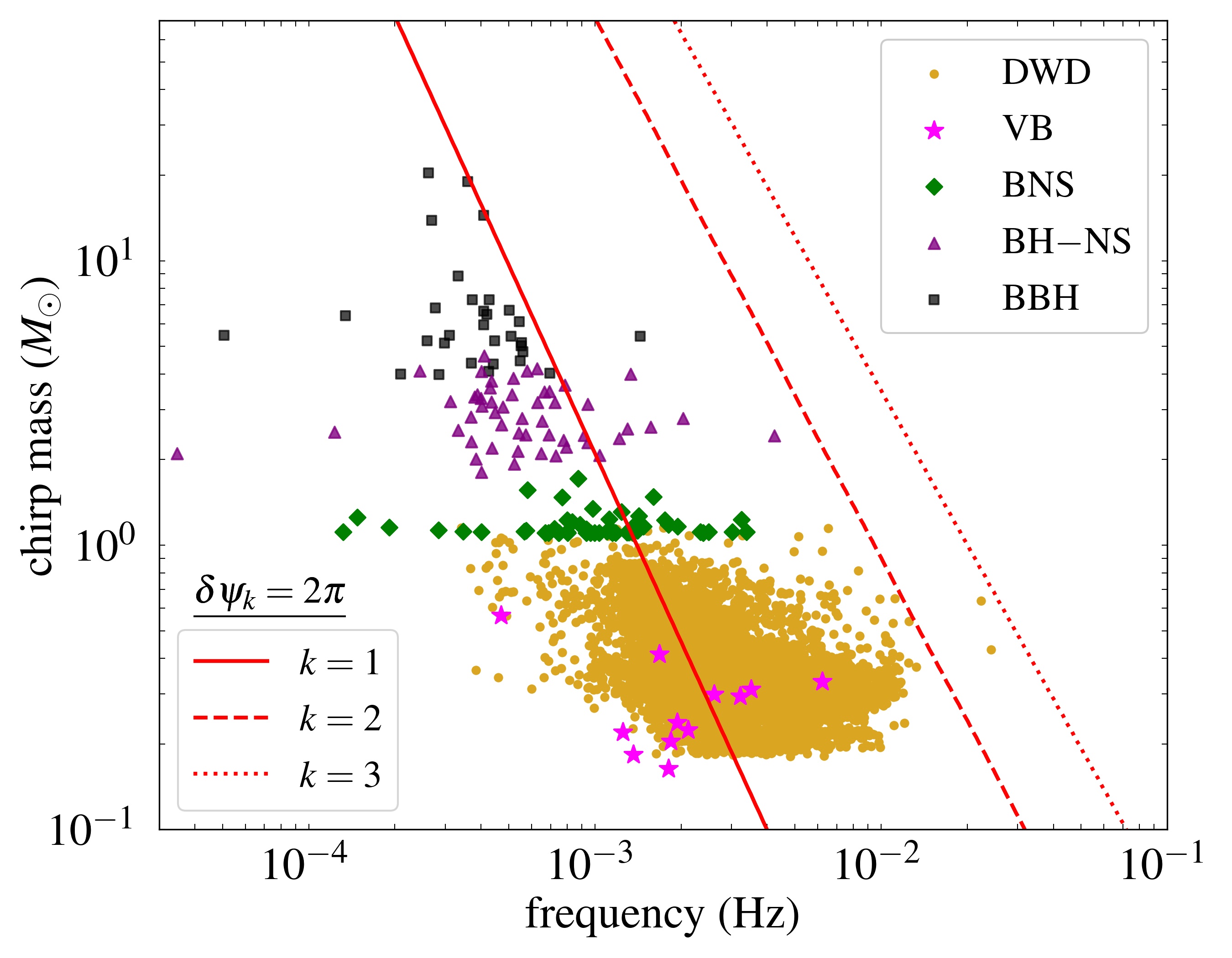}}%
    \caption{Yellow dots show the synthetic DWD binary population considered in this work in the frequency vs. chirp mass plane. Other types of compact binaries listed in the legend are verification binaries (VB), binary neutron stars (BNS), black hole--neutron star binaries (BH--NS), and binary black holes (BBH). The red lines are contours of constant dephasing $\delta\psi_k=2\pi$. The plot suggests that there should be a sizeable portion of the DWD population with measurable $\dot{f}_0$ (i.e., to the right of the $k=1$ line), while $\ddot{f}_0$ is only measurable for a few binaries (roughly, those to the right of the $k=2$ line).
    }
    \label{fig:cycles}
\end{figure} 

Figure~\ref{fig:cycles} shows the GW frequency--chirp mass plane with three contours of constant $\delta\psi_k=2\pi$ for $k=1$ (solid red line), $k=2$ (dashed), and $k=3$ (dotted). The contours correspond to the criterion of sufficient dephasing for reliable measurement of the $k$th derivative, Eq.~(\ref{eq:psik_deriv}), with sources with a smaller dephasing lying to the left of the respective line. We also overlay the synthetic population described in Section~\ref{subsec:DWDpopulation}. In addition to the DWDs and verification binaries (VB), here we also include binary neutron stars (BNS), black hole--neutron star binaries (BH--NS), and binary black holes (BBH). There is clearly a sizeable portion of the DWD population (as well as a few compact binaries) with measurable $\dot{f}_0 (k=1)$. By contrast, one can hope to measure $\ddot{f}_0 (k=2)$ only for a few DWDs, while measuring the higher derivative $f^{(3)}_0 (k=3)$ seems unlikely with LISA over a 10 yr observation period. For other types of compact binaries, $\ddot{f}_0$ may be detectable for $\sim 1$ source, whereas detecting higher derivative is unlikely. Note, however, that the dephasing criterion is approximate, so measurements may still be possible for sources with high~SNR.

To summarize: out of all resolved DWDs observable by LISA, there is likely only a few for which the derivative parameter $\ddot{f}_0$ can be measured independently of $\dot{f}_0$ from GWs alone, thereby lifting the degeneracy between the GW chirp~$\dot{f}_{\rm s0}$ and the acceleration~$a$, and enabling DWD accelerometry. We now proceed to provide quantitative estimates of the measurement uncertainties.

\subsection{Fisher uncertainties\label{subsec:FD}}

We can make the previous arguments more quantitative by calculating the apparent acceleration uncertainty for a few representative DWD sources using the Fisher information matrix.
We begin with the conservative case in which we try to determine the acceleration using only GW observations. As discussed in the previous section, this amounts to measuring the second derivative~$\ddot{f}_0$ and lifting the $\dot{f}_{\rm s0}$--$a$ degeneracy. We then consider another way to break that degeneracy, which is to ascertain one of the characteristic parameters of a DWD from EM observations. We consider both the case in which only the DWD chirp mass~$\mathcal{M}$ is assumed to be known, as well as a ``best-case scenario'' in which all parameters are known except for the acceleration.
 
For the Fisher matrix analysis, we assume a TD waveform
\beqa
h(t)&=&A\cos\psi(t)\,, \label{eq:WaveformTD}\\
\psi(t) &=& \psi_0 + 2\pi\left(f_0 t + \frac 12\dot{f}_0 t^2 + \frac 16\ddot{f}_0 t^3\right)\,, \label{eq:PhaseTD}
\eeqa
where the amplitude~$A$ is approximately constant, and $\psi_0$ is the initial phase. The GW signal is also assumed to have been averaged over the angles that characterize the sky position of the source and its orbital orientation. In what follows we carry out the calculation in the FD, where the waveform reads
\be
h_F = \left\{
\begin{array}{ll}
\displaystyle\frac 12 \frac{A}{\sqrt{\dot{f}\left[t(F)\right]}}e^{i\frac{\pi}{4}+i(\psi_0+\psi_F)}\,, & f_0\leq F\leq f(T_{\rm obs})\,, \\
{} & {} \\
0\,, & \mbox{otherwise}\,.
\end{array}
\right.
\ee
Here we use the uppercase~$F$ for the Fourier frequency, and we denote the FD waveform by the subscript~$F$. Explicit expressions for the functions~$t(F)$ and~$\psi_F$ are provided in Appendix~\ref{app:fisher}, where we also describe our numerical setup and compare three different versions of the Fisher matrix calculation (the FD, the TD, and a discretized version that emulates data processing in LISA).

The Fisher matrix~$F_{ij}$ for a source with characteristic parameters~$\theta_i$ is defined in terms of the inner product between two FD waveforms as follows:
\beqa
\left(\left.h_F^{(1)}\right|h_F^{(2)}\right) &=& \mathcal{C}^2{\rm Re}\int\limits_0^{+\infty}{\frac{h_F^{(1)}\left(h_F^{(2)}\right)^*}{S_{\rm n}(F)}\,\dd{F}}\,, \label{eq:InnerProduct}\\
F_{ij}&\equiv&\left(\left.\frac{\partial h_F}{\partial\theta_i}\right|\frac{\partial h_F}{\partial\theta_j}\right)\,, \label{eq:FisherMatrix}
\eeqa
where $S_{\rm n}$ is the noise power spectral density measured by LISA~\cite{Robson:2018ifk,2021arXiv210801167B}, and the factor~$\mathcal{C}$ absorbs all non-essential numerical constants. The inverse of the Fisher matrix (or ``correlation matrix'') is
\be
\langle\sigma_i\sigma_j\rangle = \left(F^{-1}\right)_{ij}\,,
\ee
where the angular brackets stand for the expectation value over realizations of the noise, and its diagonal components yield the uncertainties~$\sigma_i$ on each of the parameters~$\theta_i$. In what follows we also use~$\varepsilon_i$ to denote the relative uncertainty of a parameter, or equivalently, the uncertainty on the logarithm of the parameter:
\be
\varepsilon_i\equiv\frac{\sigma_i}{\theta_i}=\sigma_{\ln{\theta_i}}\,.
\ee

For the quasimonochromatic sources under consideration, it is convenient to normalize the uncertainties by the SNR~\cite{Seto:2002dz,2002ApJ...575.1030T}:
\beqa
\mbox{SNR}^2&\equiv&\left(\left.h_F\right|h_F\right) = \mathcal{C}^2\int\limits_{f_0}^{f_0 + \delta f}{\frac{\left|h_F\right|^2}{S_{\rm n}(F)}\,\dd{F}} \nonumber \\
&\approx& \left(\frac 12\mathcal{C}A\right)^2\frac{\delta f}{\dot{f}_0 S_{\rm n}(f_0)} \approx \left(\frac 12\mathcal{C}A\right)^2\frac{T_{\rm obs}}{S_{\rm n}(f_0)}\,,
\eeqa
or
\be
\mbox{SNR} = \frac 12\mathcal{C}A \sqrt{\frac{T_{\rm obs}}{S_{\rm n}(f_0)}}\,,
\ee
where $\delta f$ is the total change in frequency during the observation time, and we have neglected terms of the order of~$\delta f/f_0\sim T_{\rm obs}/\tau_{\rm c}\ll 1$. Note that the amplitude~$A$ is constant in the same approximation (see e.g. Ref.~\cite{2023arXiv231200121S}).

\begin{figure}[t]
    {\includegraphics[width=0.4\textwidth]{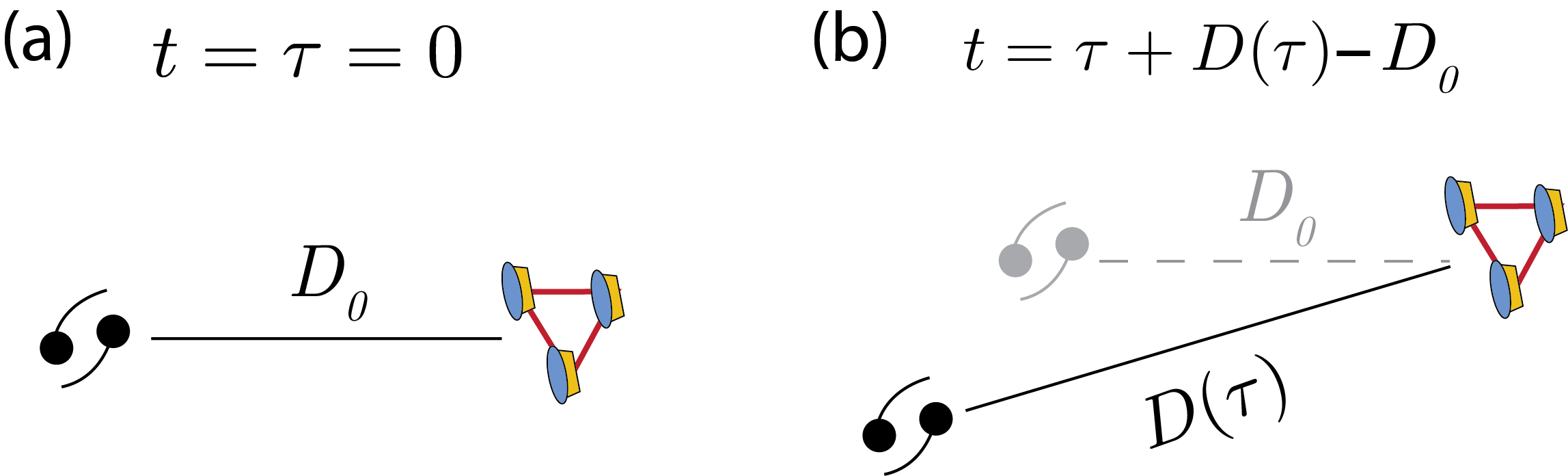}}
    \caption{(a) Beginning of DWD GW observation by LISA. (b) Time $t_s$ passes according to a clock at the source. We assume that the displacement of the source is much less than the source distance $D$ at any point during the LISA observation.}%
    \label{fig:times}%
\end{figure}

We compute Fisher matrices and the respective uncertainties for the following sets of parameters:
\begin{itemize}[leftmargin=*]
\item[(1)] $\theta_i^{(1)} = \{\ln A,\ln{f_0},\ln{\dot{f}_0}, \ln{\ddot{f}_0}, \psi_0\}$\,.
\item[(2)] $\theta_i^{(2)} = \{\ln A,\ln{f_{\rm s0}},\ln{\dot{f}_{\rm s0}}, a, \psi_0\}$\,. 

In order to obtain the change of variables between~$\theta_i^{(2)}$ and~$\theta_i^{(1)}$, we relate the observed frequency and its derivatives to their counterparts at the source. Since a constant velocity can be absorbed into redefinitions of the GW frequency and acceleration, we set the initial velocity $v_{||,0}=0$. Then, as shown in Appendix~\ref{app:shklovskii}, the line-of-sight distance to the source is
\be
D(\tau) = D_0 + \frac 12 a\tau^2\,,
\ee 
where $D_0$ is the distance at which the GW signal is first detected. Due to the light travel time effect~\cite{1952ApJ...116..211I,1959AJ.....64..149I} (see also~\cite{Anglada:2005cc}), the times at the observer and at the source are related as follows (see Fig.~\ref{fig:times}):
\beqa
t &=& \tau + D(\tau)-D_0\,, \nonumber \\
t &=& \tau + \frac 12 a\tau^2 \,. \label{eq:LTT}
\eeqa
Since the number of full-period waves emitted by the source is the same as the number received by the observer, the GW phase is invariant:
\beqa
\psi(t) &=& \psi_{\rm s}(\tau)\,, \nonumber \\
2\pi\int_0^t{f(t')\,\dd{t'}} &=& 2\pi\int_0^{\tau}{f_{\rm s}(\tau')\,\dd{\tau'}} \,,
\eeqa
where $t$ and~$\tau$ are related by Eq.~(\ref{eq:LTT}). Differentiating both sides w.r.t. the observer's time~$t$, we obtain~\cite{Bonvin:2016qxr,Meiron:2016ipr,Tamanini:2019usx,Robson:2018svj}:
\be
f(t) = \frac{f_{\rm s}(\tau)}{\dd{t}/\dd\tau}=\frac{f_{\rm s}\left[\tau(t)\right]}{1 + a\tau(t)}\,,
\ee
where $\tau(t)$ is the inverse of~$t(\tau)$. Therefore,
\beqa
f_0 &=& f_{\rm s0}\,, \\
\dot{f}_0 &=&\dot{f}_{\rm s0}-af_{\rm s0}\,, \label{eq:fdota-conversion-1}\\
\ddot{f}_0 &=& \ddot{f}_{\rm s0} - 3a\dot{f}_{\rm s0} + \mathcal{O}(a^2 f_{\rm s0}) \nonumber \\
&\approx&\frac{11}{3}\frac{\dot{f}_{\rm s0}^2}{f_{\rm s0}}- 3a\dot{f}_{\rm s0}\,, \label{eq:fdota-conversion-2}
\eeqa
where we used Eq.~(\ref{eq:exact-fdots}) to relate the derivatives at the source. We can estimate the relative correction due to acceleration as $af_{\rm s0}/\dot{f}_{\rm s0}\sim a\tau_{\rm c} \ll 1$. Note that an overdot means that the frequencies are differentiated with respect to their own times: $t$ for the observer and~$\tau$ for the source.

The final change of variable reads:
\beqa
\ln{f_0} &=& \ln{f_{\rm s0}}\,, \label{eq:theta12-1}\\
\ln{\dot{f}_0} &=&\ln{\dot{f}_{\rm s0}}-ae^{\ln{f_{\rm s0}}-\ln{\dot{f}_{\rm s0}}}\,, \label{eq:theta12-2}\\
\ln{\ddot{f}_0} &=&2\ln{\dot{f}_{\rm s0}}-\ln{f_{\rm s0}}-\frac{9}{11}ae^{\ln{f_{\rm s0}}-\ln{\dot{f}_{\rm s0}}}\,, \label{eq:theta12-3}
\eeqa
where we have omitted a constant that does not contribute to the Jacobian between~$\theta_i^{(2)}$ and~$\theta_i^{(1)}$.

\item[(3)] $\theta^{(3)}_i=\{\ln\mathcal{M},\ln D,\ln{\tau_{\rm c}},a,\psi_{\rm c}/\beta\}$\,, with 
\beqa 
\psi_{\rm c}&\equiv&\psi_0 + \frac{1}{16}(\pi\mathcal{M}{f_{\rm s0}})^{-5/3}\,,  \\
\beta&\equiv& 2\pi f_{\rm s0} \tau_{\rm c}=\frac{5}{128}(\pi\mathcal{M}{f_{\rm s0}})^{-5/3}\,,
\eeqa 
where $\psi_{\rm c}$ is the coalescence phase, and the factor $\beta$ is introduced for numerical convenience. Using Eq.~(\ref{eq:coalesc-time}) and $A\approx \mathcal{M}^{5/3}f_{\rm s0}^{2/3}/D$, the change of variables between~$\theta_i^{(3)}$ and~$\theta_i^{(2)}$ (to within an additive constant) reads
\beqa
\ln A &=& \frac 54\ln\mathcal{M} - \frac 14\ln\tau_{\rm c} - \ln D\,, \label{eq:theta23-1} \\
\ln f_{\rm s0} &=& -\frac 58\ln\mathcal{M} - \frac 38\ln\tau_{\rm c}\,, \label{eq:theta23-2} \\
\ln \dot{f}_{\rm s0} &=& -\frac 58\ln\mathcal{M} - \frac{11}{8}\ln\tau_{\rm c}\,, \label{eq:theta23-3} \\
\psi_0 &=& \beta\left(\frac{\psi_{\rm c}}{\beta}-\frac 85\right)\,. \label{eq:theta23-4}
\eeqa
\end{itemize}

\begin{table}
\caption{Fisher uncertainties (diagonal) and correlation coefficients (off-diagonal) for two sets of characteristic DWD GW parameters, respectively: $\theta^{(1)}_i=\left\{\ln A,\ln f_0,\ln\dot{f}_0,\ln\ddot{f}_0,\psi_0\right\}$ (top) and $\theta^{(2)}_i=\left\{\ln A,\ln{f_{\rm s0}},\ln{\dot{f}_{\rm s0}}, a,\psi_0\right\}$ (bottom). We assume $\mbox{SNR}=1000$, with $\mathcal{M}=0.44M_\odot$, $f_0=2$~mHz, $T_{\rm obs}=10$~yr.\label{tab:consistency-fddot}}
\begin{tabular}{c|ccccc}\hline\hline\\ 
$\theta^{(1)}_i$ & $\varepsilon_A$  & $\varepsilon_{f_0}$  & $\varepsilon_{\dot{f}_0}$ & $\varepsilon_{\ddot{f}_0}$ & $\sigma_{\psi_0}$ \\
\\ \hline \\
$\varepsilon_A$ & $0.001^*$ & $0^*$ & $0^*$ & $0^*$ & $0^*$   \\
$\varepsilon_{f_0}$ & {} & $8.741\times 10^{-9}$ & $-0.97$ & 0.92 & $-0.87$ \\
$\varepsilon_{\dot{f}_0}$ & {} & {} & $\mathbf{0.01375}$ & $\mathbf{-0.99}$ & $0.75$ \\
$\varepsilon_{\ddot{f}_0}$ & {} & {} & {} & $\mathbf{2504}$ & $-0.66$ \\
$\sigma_{\psi_0}$ & {} & {} & {} & {} & 0.004 \\
\hline\hline\\
$\theta^{(2)}_i$ & $\varepsilon_A$  & $\varepsilon_{f_{\rm s0}}$  & $\varepsilon_{\dot{f}_{\rm s0}}$ & $\sigma_a~[\mathrm{m/s^2}]$ & $\sigma_{\psi_0}$ \\
\\ \hline \\
$\varepsilon_A$ & $0.001^*$ & $0^*$ & $0^*$ & $0^*$ & $0^*$   \\
$\varepsilon_{f_{\rm s0}}$ & {} & $8.741\times 10^{-9}$ & 0.92 &  0.92 & $-0.87$ \\
$\varepsilon_{\dot{f}_{\rm s0}}$ & {} & {} & $\mathbf{2119}$ &  $\mathbf{1}$ & $-0.66$ \\
$\sigma_a~[\mathrm{m/s^2}]$ & {} & {} & {} & $\mathbf{0.006}$  & $-0.66$ \\
$\sigma_{\psi_0}$ & {} & {} & {} & {} & 0.004 \\
\hline
\end{tabular}
\end{table} 

\noindent
{\bf \textit{GWs alone}}. Table~\ref{tab:consistency-fddot} shows the results of a FD calculation of Fisher uncertainties for the sets of parameters $\theta^{(1)}_i$ (top) and $\theta^{(2)}_i$ (bottom), assuming no additional information beyond the measured GW waveform. Recall that the first set includes the observed derivatives~$\dot{f}_0$ and~$\ddot{f}_0$, while in the second set they are converted to the source value~$\dot{f}_{\rm s0}$ and the acceleration~$a$. Uncertainties on these parameter sets are related by the Jacobian for the change of variables given in Eqs.~(\ref{eq:theta12-1})--(\ref{eq:theta12-3}), and the calculation is performed for a loud quasimonochromatic source with $\mathcal{M}=0.44M_\odot$, $f_0=2$~mHz, $T_{\rm obs}=10\text{ yr}$, and $\mbox{SNR}=1000$. Comparing the values highlighted in bold makes emphasizes quantitatively that the double derivative~$\ddot{f}_0$ is difficult to measure, resulting in a high uncertainty for the apparent acceleration. To reiterate: if we do not include~$\ddot{f}_0$ in the set of Fisher matrix parameters, there is complete degeneracy between $\dot{f}_{\rm s0}$ and~$a$, because we can only measure their combination~$\dot{f}_0$, Eq.~(\ref{eq:fdota-conversion-1}). The inclusion of~$\ddot{f}_0$ introduces a linearly independent combination, Eq.~(\ref{eq:fdota-conversion-2}), breaking the degeneracy in principle, but this additional parameter is poorly determined for realistic DWD parameters: while~$\dot{f}_0$ is measured well ($\approx 1\%$), there are large uncertainties for $\dot{f}_{\rm s0}$ and~$a$ from the inversion of Eqs.~(\ref{eq:fdota-conversion-1}) and~(\ref{eq:fdota-conversion-2}).

\noindent
{\bf \textit{GWs + EM observations}}. The situation improves significantly if an independent measurement of either the DWD chirp mass~$\mathcal{M}$ or distance~$D$ is available from EM observations. If the chirp mass is known, it relates the time derivative $\dot{f}_{\rm s0}$ to the frequency~$f_{\rm s0}=f_0$, which is always measured very well. A reasonably good measurement of $\dot{f}_0$ then leads to a dramatic improvement in~$\sigma_a$. If, instead, the distance is known, the chirp mass can be inferred from the amplitude~$A$, whose uncertainty is relatively low (see Table~\ref{tab:consistency-fddot}). Knowing the chirp mass again makes it possible to infer the acceleration with low uncertainty (i.e., small $\sigma_a$).

Tables~\ref{tab:TDwithMapprox} and~\ref{tab:TDwithoutMapprox} show the apparent acceleration uncertainty~$\sigma_a$ at~$\mbox{SNR}=1000$ when the chirp mass~$\mathcal{M}$ is either included or excluded from the vector of parameters~$\theta_i^{(3)}$. For the fiducial values of the parameters, we choose the combinations $\mathcal{M}=0.44M_\odot$, $f_0=2$~mHz  and~$\mathcal{M}=M_\odot$, $f_0=5$~mHz, while for the observation time we use either $T_{\rm obs}=4$~yr or $T_{\rm obs}=10$~yr. Note that the uncertainty improves $\propto T_{\rm obs}^{-2}$ (at fixed SNR, i.e., not accounting for the scaling $\text{SNR}^{-1}\propto T_{\rm obs}^{-1/2}$) as one may expect from dimensional considerations. We also find that the uncertainty barely depends on the exact value of~$a$ as long as Galactic accelerations are concerned, which is why we assume $a=0$ in this calculation.

\begin{table}
\caption{Apparent acceleration uncertainty~$\sigma_{\rm a}~[\mathrm{m/s^2}]$ for a set of characteristic DWD GW parameters with the chirp mass \textit{included}: $\theta^{(3)}_i=\{\ln\mathcal{M},\ln D,\ln{\tau_{\rm c}},\psi_{\rm c}/\psi_0,a\}$. Fiducial values of the chirp mass, GW frequency, and observation time are shown; we assume $\mbox{SNR}=1000$ and~$\psi_0=0$.\label{tab:TDwithMapprox}}%
\begin{center}
\begin{tabular}{c|c|c}\hline\hline
\begin{tabular}{@{}c@{}}Time domain,\\$\mathcal{M}\in \theta^{(3)}_i$ \end{tabular} & \begin{tabular}{@{}c@{}}$\mathcal{M}=0.44M_\odot$ \\ $f_0=2$~mHz\end{tabular} & \begin{tabular}{@{}c@{}}$\mathcal{M}=1M_\odot$ \\ $f_0=5$~mHz\end{tabular} \\ \hline\hline
{} & {} & {} \\
$T_{\rm obs}=4$~yr & $6.73\times 10^{-2}$ & $5.47\times 10^{-4}$ \\
{} & {} & {}\\
$T_{\rm obs}=10$~yr & $4.30\times 10^{-3}$ & $2.04\times 10^{-5}$ \\
{} & {} & {} \\
\hline
\end{tabular}
\end{center}
\end{table}

\begin{table}
\caption{Apparent acceleration uncertainty~$\sigma_{\rm a}~[\mathrm{m/s^2}]$ for a set of characteristic DWD GW parameters with the chirp mass \textit{excluded}: $\theta^{(3)}_i$ = $\{\ln D,\ln{\tau_{\rm c}},\psi_{\rm c}/\psi_0,a\}$. Fiducial values of the chirp mass, GW frequency, and observation time are shown; we assume $\mbox{SNR}=1000$ and~$\psi_0=0$.\label{tab:TDwithoutMapprox}}%
\begin{center}
\begin{tabular}{c|c|c}\hline\hline
\begin{tabular}{@{}c@{}}Time domain,\\$\theta^{(3)}_i\,\backslash\,\{\mathcal{M}\}$ \end{tabular} & \begin{tabular}{@{}c@{}}$\mathcal{M}=0.44M_\odot$ \\ $f_0=2$~mHz\end{tabular} & \begin{tabular}{@{}c@{}}$\mathcal{M}=1M_\odot$ \\ $f_0=5$~mHz\end{tabular} \\ \hline\hline
{} & {} & {} \\
$T_{\rm obs}=4$~yr & $2.84\times 10^{-8}$ & $1.14\times 10^{-8}$ \\
{} & {} & {}\\
$T_{\rm obs}=10$~yr & $4.54\times 10^{-9}$ & $1.82\times 10^{-9}$ \\
{} & {} & {} \\
\hline
\end{tabular}
\end{center}
\end{table}

\noindent
{\bf \textit{A best-case scenario.}} In an ideal situation, all parameters of the source would be known exactly, except for the acceleration. The TD Fisher ``matrix'' (see Appendix~\ref{subsec:app:implement}) then consists of a single element:
\beqa
F_{\rm aa} &=& \mathcal{C}^2\int\limits_0^{T_{\rm obs}}{\frac{\left(\partial_{a}h\right)^2}{S_n[f(t)]}\,\dd{t}} \approx \frac{(\mathcal{C}A)^2}{2S_n(f_0)}(\pi f_{0})^2 \int\limits_0^{T_{\rm obs}}{\dd{t}\,t^4\sin^2{\psi}} \nonumber \\
&\approx& \mbox{SNR}^2\times\frac{(\pi f_{0})^2}{T_{\rm obs}}\int\limits_0^{T_{\rm obs}}{\dd{t}\,t^4(1-\cos{2{\psi}})} \nonumber \\
&\approx& \mbox{SNR}^2\,T_{\rm obs}^4\times\frac{(\pi f_{0})^2}{5}\,,
\eeqa
where we neglect $t/\tau_{\rm c}\lesssim T_{\rm obs}/\tau_{\rm c}\ll 1$ and $\mathcal{O}[(f_0T_{\rm obs})^{-1}]\ll 1$ terms. Then, the acceleration uncertainty is given by:
\beqa
\sigma_{\rm a} &=& \frac{1}{\sqrt{F_{\rm aa}}} = \frac{1}{\mbox{SNR}}\frac{1}{T_{\rm obs}^2}\frac{\sqrt{5}}{\pi f_{0}} \nonumber \\
&\approx& 10^{-9}\;\mbox{m/s}^2\,\left(\frac{\mbox{SNR}}{1000}\right)^{-1}\left(\frac{T_{\rm obs}}{10\;\mbox{yr}}\right)^{-2}\left(\frac{f_0}{2\;\mbox{mHz}}\right)^{-1}\,. \nonumber \\
&{}& \label{eq:acc_lower}
\eeqa
This estimate gives a lower limit on the uncertainty. Indeed, as we include more parameters in the Fisher matrix we introduce more correlations, which leads to higher uncertainties on the individual parameters.

As the results of this section suggest, measuring the Galactic acceleration with GWs alone appears to be unlikely with LISA. The typical values of the uncertainty in this case (see Table~\ref{tab:TDwithMapprox}) are orders of magnitudes higher than the typical acceleration: see Fig.~\ref{fig:accel}. When the chirp mass is known, the uncertainty improves dramatically, although it is still an order of magnitude worse than the typical magnitude of Galactic acceleration.

There is still some hope: since we are interested in large-scale Galactic accelerations, the accelerations of the individual DWDs in the population must be correlated, which should allow improvement of the overall uncertainty. We discuss this possibility in the next section.

\begin{figure*}
    \centering
    \includegraphics[width=0.495\textwidth]{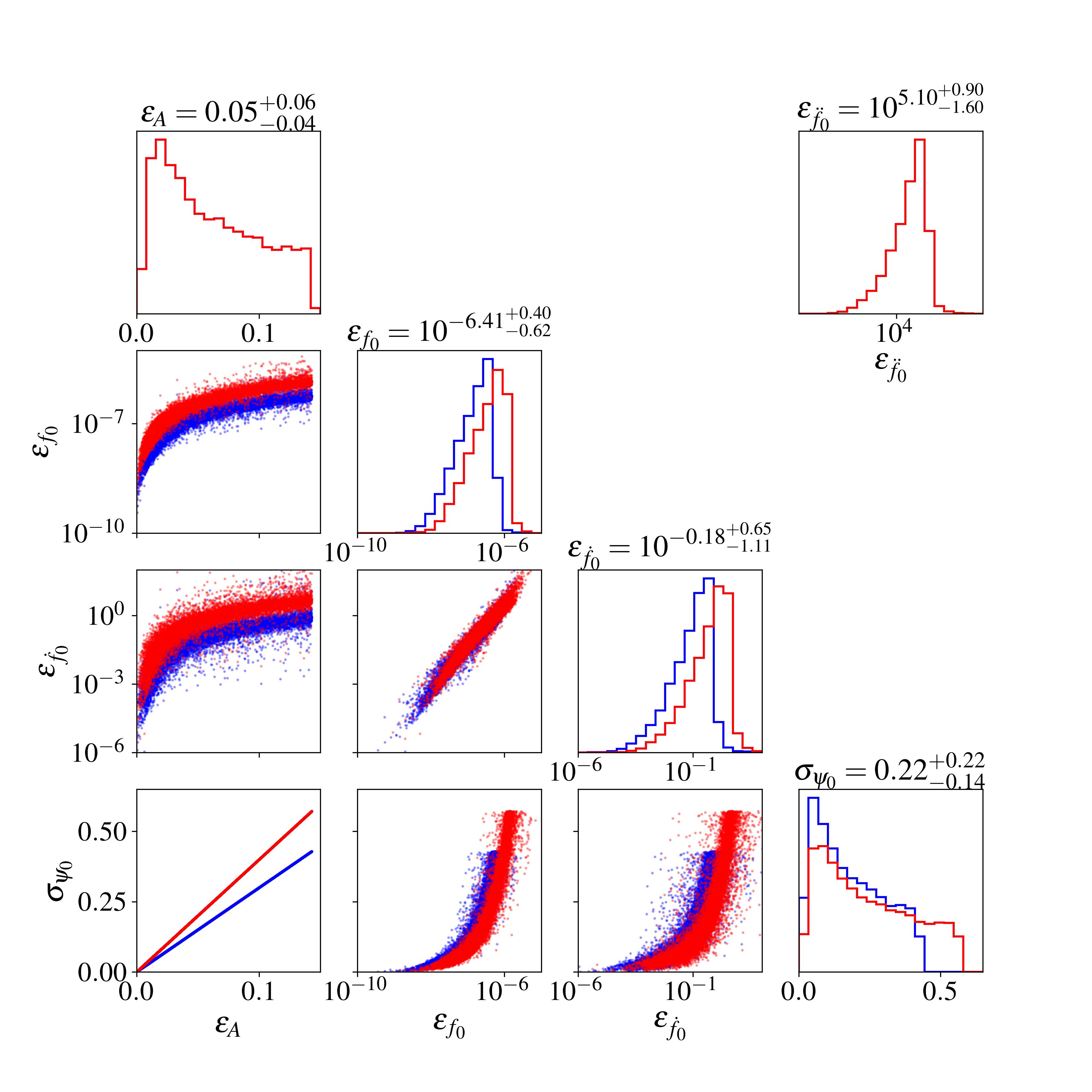}
    \includegraphics[width=0.495\textwidth]{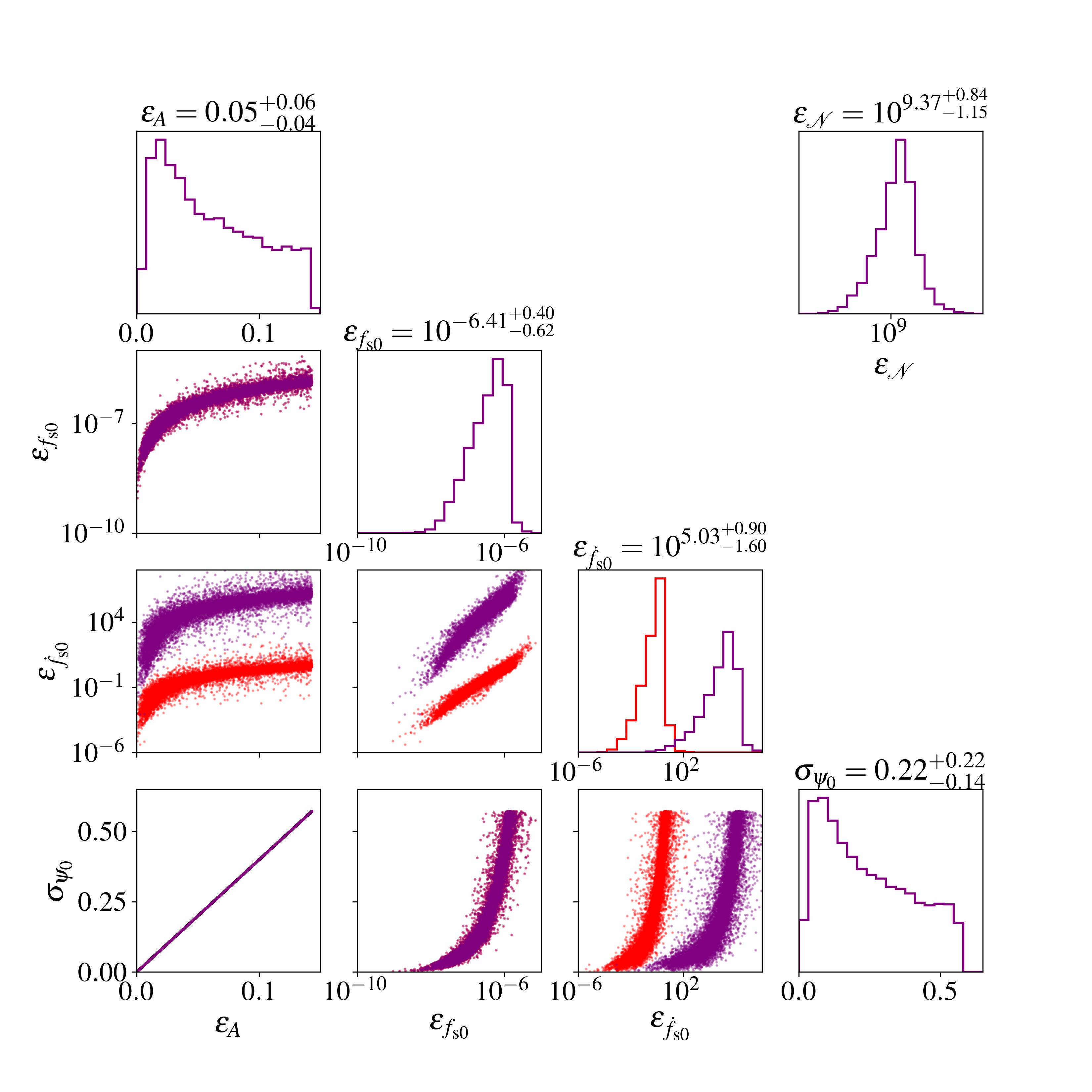}
    \caption{Measurement uncertainties for a synthetic population of Galactic DWDs (model FZ from Ref.~\cite{Thiele:2021yyb}) using GWs alone. \textit{Left}: The second time derivative of the GW frequency~$\ddot{f}_0$ is either included (red) or excluded (blue) from the set of parameters in the Fisher matrix calculation. Here, numbers quoted at the top of plots refer to the ``red'' case. \textit{Right}: The parameters include either the observed first and second time derivatives of the GW frequency (red, same points as in the left panel), or the source GW frequency~$\dot{f}_{\rm s0}$ and the global parameter~$\mathcal{N}$ (purple), which is related to the Galactic gravitational potential as in Eq.~(\ref{eq:overall_mass_normalization}). Here, numbers quoted at the top of the plots refer to the ``purple'' case.
    \label{fig:population_errors_GW}}%
\end{figure*}

\section{Galactic accelerometry with gravitational waves \label{sec:accelerometry}}

In this section we present the distribution of measurement uncertainties for the synthetic DWD population described in Section~\ref{subsec:DWDpopulation} above. We then discuss the prospects of measuring their acceleration by leveraging the large number of Galactic DWDs and using additional information from EM observations. To model the correlation between the accelerations of individual DWDs, we introduce an overall normalization factor, $\mathcal{N}$, of the Galactic gravitational potential. Effectively, the parameter $\mathcal{N}$ encodes a linear scaling with the total mass of the Milky Way:
\begin{equation}\label{eq:overall_mass_normalization}
    \Phi = \mathcal{N}\Phi_{\rm model}\,,
\end{equation}
where $\Phi_{\rm model}$ is defined in Eqs.~\eqref{eq:Phimodel} and \eqref{eq:Phi_comps}.
Since $\vec{a}=-\nabla\Phi$ and $v^2=R\,\partial\Phi/\partial R$ (see Section\,\ref{sec:gal_accel_A}), normalizing the gravitational potential also normalizes the apparent acceleration as $a=\mathcal{N}a_{\rm model}$.

The GW signals from multiple resolvable DWDs located at different distances from the Galactic center can be used to collectively pin down the parameter $\mathcal{N}$ as follows. Consider a number~$n$ of DWDs that are used to determine~$\mathcal{N}$. Then, if each DWD source is described by~$k$ parameters (chirp mass, distance, etc.), we can calculate a large Fisher matrix of size $(nk+1)\times(nk+1)$ for the global parameter vector $\left(\mathcal{N},\boldsymbol{\theta}_1,\boldsymbol{\theta}_2\,,\ldots\,,\boldsymbol{\theta}_n\right)$, where $\boldsymbol{\theta}_i$ is the parameter vector of the $i$th source. If we assume that the narrow GW frequency bands swept by any two DWDs do not overlap, the Fisher matrix reduces to $n$ blocks of individual $k\times k$ Fisher matrices, with the first row and column (of ``size'' $1\times (nk+1)$ and $(nk+1)\times 1$, respectively) encoding correlations between the parameter~$\mathcal{N}$ and the individual sets of parameters~$\boldsymbol{\theta}_i$ $(i=1\,,\ldots\,, n)$. It then follows that the reciprocals of the individual uncertainties add in quadrature to provide an improvement that scales with $1/\sqrt{n}$. The parameter $\mathcal{N}$ is considered measurable if the fractional error
$\varepsilon_\mathcal{N}=\sigma_{\mathcal{N}}/\mathcal{N}<1$, where $\mathcal{N}=1$ corresponds to a fiducial model of the Milky Way such that $\Phi=\Phi_\mathrm{model}$.
Conversely, if $\varepsilon_\mathcal{N}>1$, the measurement only places an upper bound on the parameter~$\mathcal{N}$.

We now proceed to present the results of our calculation; the details of our fast implementation can be found in Appendix~\ref{subsec:app:implement}.

\begin{figure}
    \centering
    \includegraphics[width=0.495\textwidth]{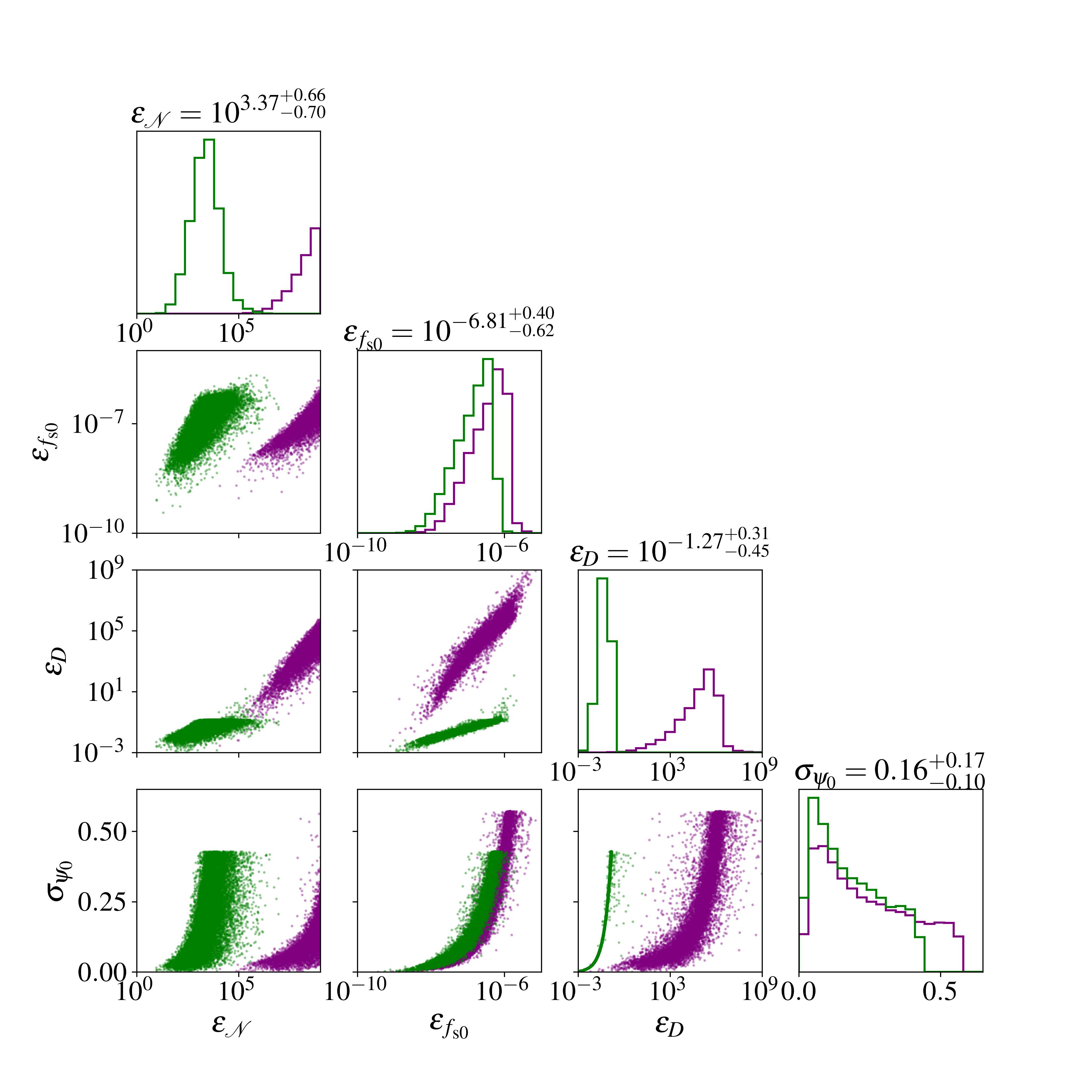}
    \caption{Comparison of the measurement uncertainties on the parameters~$(\mathcal{N},\mathcal{M},D,f_{\rm s0},\psi_0)$ when the chirp mass~$\mathcal{M}$ is included (purple) or excluded (green) from the parameter set. The numbers quoted at the top of the plots refer to the ``green'' case. Uncertainties on~$\mathcal{M}$ are not shown even when this parameter is included, because in general it cannot be determined. The synthetic population of DWDs is the same as in Fig.~\ref{fig:population_errors_GW} above (i.e., model FZ from Ref.~\cite{Thiele:2021yyb}).
    \label{fig:population_errors_EM}}%
\end{figure}

\subsection{Distribution of measurement uncertainties}

\noindent
{\bf \textit{GWs alone.}} In Fig.~\ref{fig:population_errors_GW} we show the distribution of measurement uncertainties for the synthetic DWD population. The left panel illustrates how the uncertanties change if the set of 4~parameters~$(A,f_0,\dot{f}_0,\psi_0)$, shown in blue, is extended to include~$\ddot{f}_0$ (red). The right panel is obtained by the change of variables defined by Eqs.~(\ref{eq:fdota-conversion-1})--(\ref{eq:fdota-conversion-2}) from the set of parameters $(f_0,\dot{f}_0,\ddot{f}_0)$, shown in red as before, to $(f_{\rm s0},\dot{f}_{\rm s0},\mathcal{N})$, now shown in purple; it illustrates the effect of including $\mathcal{N}=a/a_0$ as an extra parameter in the set~$(A,f_{\rm s0},\dot{f}_{\rm s0},\psi_0)$. Here $a_0$ stands for the acceleration in a fiducial Milky Way model. The relative uncertainties quoted above the top-right subplots correspond to the extended parameter sets (distributions in purple).

As is evident from the left panel, while the parameter $\ddot{f}_0$ is far from being measurable, its inclusion in the Fisher matrix has little effect on the inference of other parameters. The parameters $A$ and $f_0$ are well measured for most of the sources (at the level of $\varepsilon_A\lesssim 10\%$ and $\varepsilon_{f_0}\lesssim 10^{-6}$, respectively). The parameter $\dot{f}_0$ is only measurable at the level of $\varepsilon_{\dot{f}_0}\lesssim 10\%$ for some of the sources. However, as we change variables from  $(f_0,\dot{f}_0,\ddot{f}_0)$ to $(f_{\rm s0},\dot{f}_{\rm s0},\mathcal{N})$, the source frequency $\dot{f}_{\rm s0}$ and normalization factor $\mathcal{N}$ become essentially unmeasurable. By contrast, the observed time derivative $\dot{f}_0$ can be measured at the $10\%$ level or better for a significant fraction of the sources. This is once again a manifestation of the degeneracy between~$f_{\rm s0}$ and~$a$: the linear combination $\dot{f}_0=\dot{f}_{\rm s0}-\mathcal{N}a_0f_{\rm s0}$ is measured more precisely than the individual parameters of interest. 

\noindent
{\bf \textit{GWs + EM observations.}} In Fig.~\ref{fig:population_errors_EM} we show a comparison of the distribution of measurement uncertainties for the same synthetic DWD population, for the case of five parameters $(\mathcal{N},\mathcal{M},D,f_{\rm s0},\psi_0)$ (purple) and the case of four parameters in which the chirp mass~$\mathcal{M}$ is known (from EM observations) but has been excluded (green) from the Fisher matrix calculation. Note that the uncertainty in chirp mass is not shown, and the five-parameter case is the same as in right panel of Fig.~\ref{fig:population_errors_GW} (red), but with the parameters $(A,\dot{f}_{\rm s0})$ traded for~$(\mathcal{M},D)$. 

Clearly, as the number of parameters is reduced, the uncertainties improve dramatically. When the chirp mass is included in the parameter set to be estimated, neither $\mathcal{N}$ nor $\mathcal{M}$ can be determined with significance, which is another manifestation of the degeneracy between $\dot{f}_{\rm s0}$ and $a$. Since $\dot{f}_{\rm s0}$ is not easy to measure, there is hardly a way to determine $\mathcal{M}$ from GWs alone. This translates into a large uncertainty in the source distance~$D$, even though the amplitude $A\propto\mathcal{M}^{5/3}f_{\rm s0}^{2/3}/D$ is known to within~$\approx 10\%$ (the GW frequency $f_{\rm s0}$ itself can always be measured). However, if we know~$\mathcal{M}$ (for example, from optical observations), those degeneracies are lifted.    

The fact that the parameter~$\mathcal{N}$ gives rise to correlations among the accelerations of individual DWDs does not remedy the problem. The uncertainty on the normalization parameter is~$\varepsilon_\mathcal{N}\sim 10^9$ (see Fig.~\ref{fig:population_errors_GW}, right panel), and we can only expect an improvement by a factor of order $\sqrt{n}\sim 10^2$ times if we take the whole LISA DWD population into account. Given that determining the Galactic acceleration from GWs alone does not appear to be feasible, the role of combined GW--EM observations becomes crucial.

\subsection{Prospects for multimessenger Galactic accelerometry}

Additional information inferred from EM measurements can help to break degeneracies in DWD parameters inferred through GWs only.  For example, depending on the inclination angle $\iota$ of the source, EM data can improve the estimate on the GW amplitude. It is also possible to have precise EM measurements of DWD GW chirp rate $\dot{f}_0$, e.g., via photometric time variability or spectroscopic radial-velocity shift measurements~\cite{2012A&A...544A.153S,Shah:2013ema}. In addition, astrometric measurements of source distance $D$ (e.g., from Gaia), in combination with GW amplitudes $A$ and frequencies $f_0$ observed by LISA, may enable inference of DWD chirp mass $\mathcal{M}$~\cite{Kupfer:2023nqx}. 
 
\begin{figure*}[htpb!]
    \includegraphics[width=\columnwidth]{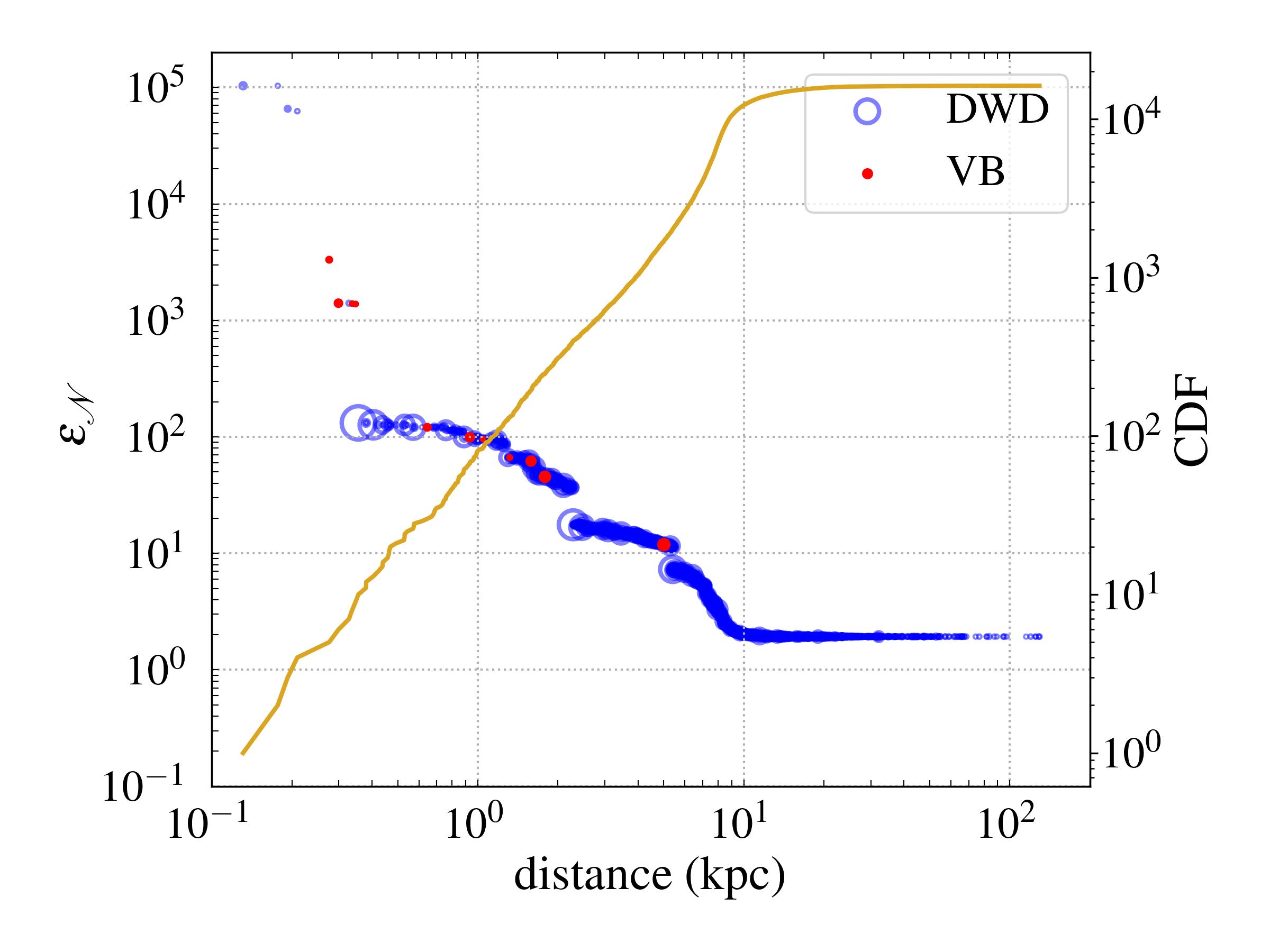}
    \includegraphics[width=\columnwidth]{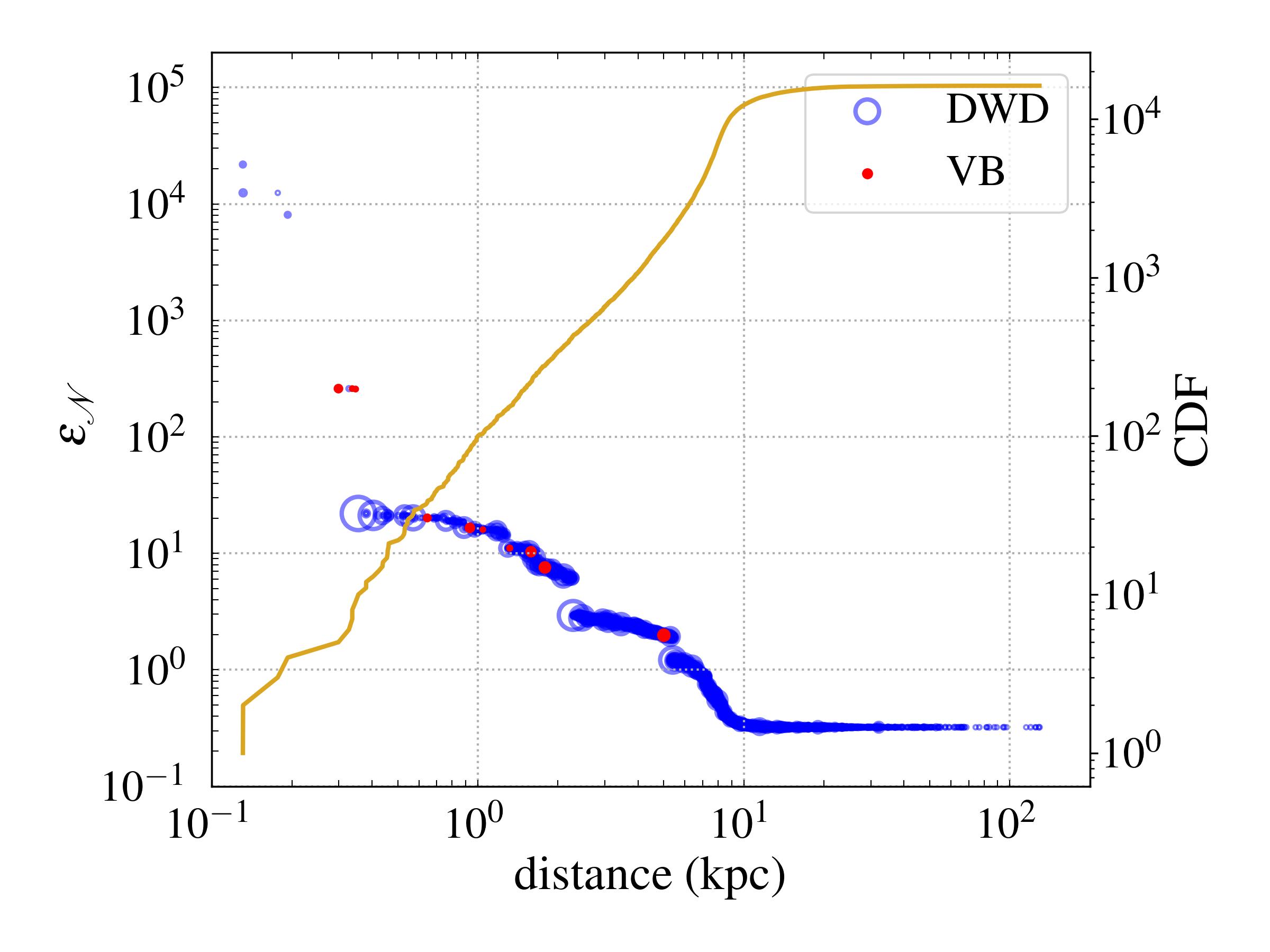}
    \caption{\textit{Left panel:} only the chirp mass $\mathcal{M}$ is determined through EM observations. \textit{Right panel:} all parameters (other than normalized acceleration $\mathcal{N}$) are determined through EM observations: i.e., $\{\mathcal{M},D,f_{\rm s0},\psi_0\}$. In each plot, the vertical position of a given circle or dot (as measured by the left vertical axis) indicates the relative measurement uncertainty on ~$\mathcal{N}$, $\varepsilon_{\mathcal{N}}\equiv\sigma_{\mathcal{N}}/\mathcal{N}$, assuming that all DWDs within a certain distance from the Sun (as given by the position of the circle or dot on the horizontal axis) were used to infer the acceleration; the orange line and the right axis show the cumulative distribution function (CDF) of the DWDs within that distance. Recall that $\mathcal{N}=a/a_0$, where $a_0$ is the acceleration in the fiducial model of the Milky Way: see also Eq.~(\ref{eq:overall_mass_normalization}). Individual sources are marked in blue (for all sources in the catalog, i.e., those detectable by LISA) or red (for verification binaries). The sizes of the circles or dots are proportional to the source SNR. Although the distance is not the only factor determining the GW SNR of a source, we observe a general trend of SNR falling with distance.
    \label{fig:orderedByDistance}}%
\end{figure*}

EM counterparts of GW sources may be identified before, during, or after the LISA observation period. Prior to the start of LISA observation, detached DWDs can be discovered in high-cadence photometric surveys as periodically variable sources, if they are sufficiently edge-on. Mass-transferring and tidal-deforming DWDs will in general show photometric time variability; however we only consider detached DWDs in this work.  With new data from current and upcoming astrometric~\cite{Gaia:2016zol}, photometric~\cite{2009arXiv0912.0201L,Gaia:2016zol,2017arXiv171103234K,2012arXiv1206.3569Z,2019Msngr.175....3D,2014SPIE.9147E..0LD}, and spectroscopic~\cite{Gaia:2016zol,2018PASP..130f4505T, Dyer:2020dmo} surveys, the number of verification binaries is expected to grow to a significantly larger number by the time of LISA launch. The increased number of verification binaries will also increase the likelihood of finding several high-SNR ($\text{SNR}\sim 100-1000$) GW sources in the collection. Furthermore, the DWD localization provided by GW measurements, if accurate enough, may allow identification of the EM counterparts through EM follow-up observations after LISA's launch. The analysis of Ref.~\cite{Littenberg:2012vs} shows that up to several hundreds of EM counterparts can be found in this way. 
We suggest that an efficient approach involves GW localization followed by targeted EM follow-up, focusing on
high-SNR, well-localized sources (since the highest SNR sources dominate the accelerometry analysis, as discussed below). According to Ref.\,\cite{Takahashi:2002ky}, the GW localization volume for high-SNR ($\gtrsim$ 100) DWD sources is typically $\lesssim$ 10 pc$^3$. The expected number of enclosed DWDs within this localization volume is less than one, based on their number density \cite{2018MNRAS.480.3942H}. Therefore, such a localization range offers a feasible target for coordinated EM observations. Further characterization of this proposed method is part of ongoing work to support future multimessenger observations. 

We now forecast how the relative error in the normalization of the Galactic potential $\varepsilon_{\mathcal{N}}=\sigma_{\mathcal{N}}/\mathcal{N}$ improves as surveys of the EM counterparts of LISA sources become more complete. For this estimate, we use the synthetic DWD population discussed in Sec.\,\ref{subsec:DWDpopulation}. LISA sources located closer to the Sun will presumably have EM counterparts discovered earlier in magnitude-limited surveys. We assume for simplicity that at any given time the EM surveys have a certain coverage distance; and, as such, there is a spherical volume centered at the Sun (with radius given by this coverage distance) within which \textit{all} LISA sources have their EM counterparts identified. 

Figure~\ref{fig:orderedByDistance} shows how $\varepsilon_{\mathcal{N}}$ improves as a function of the coverage distance, assuming that EM measurements determine either (1) $\mathcal{M}$ only (left panel), or (2) all parameters except $\mathcal{N}$ (right panel), i.e., $\{\mathcal{M},D,f_{\rm s0},\psi_0\}$. In both cases we see that, as the coverage distance grows, the relative error $\varepsilon_\mathcal{N}$ decreases, owing to the growing number of EM counterparts (recall that the reciprocals of individual uncertainties add in quadrature: see Section~\ref{sec:accelerometry}). The gradual decrease in the cumulative error is occasionally interrupted by sharp drops whenever a DWD with very high SNR (up to $\sim 1000$) is included in the coverage. In case (1), we find that the relative error asymptotes to $\epsilon_{N}\approx 2$ at coverage distance $\gtrsim 10\text{ kpc}$. This suggests that even after all of the EM counterparts have been found, the parameter~$\mathcal N$ is not quite measureable (but almost so). On the other hand, in case (2), the estimated value of $\epsilon_{\mathcal{N}}$ is about one order magnitude lower than in case (1), and the parameter~$\mathcal{N}$ is measurable once there is identification of $\sim 10^3$ EM counterparts within a few kpc. This is an encouraging result, as the current catalog of verification binaries already includes DWDs with distances of a few kpc~\cite{Kupfer:2023nqx}.

Given this analysis, using LISA observations for Galactic accelerometry appears feasible as part of a multimessenger effort. Another avenue for improvement is to take into account the correlation between the acceleration and location of a GW source in the Galaxy, encoded in its distance and sky position; we leave this more sophisticated analysis to future work. Also, our forecast is based on current expectations for LISA sensitivity and observation period. If the level of noise in LISA is lower than assumed here, which is possible given the encouraging results from LISA Pathfinder~\cite{Armano:2016bkm}, this will further reduce measurement uncertainties. We also note that using synthetic DWD populations models different from the one discussed in Sec.\,\ref{subsec:DWDpopulation} would yield slightly different results. For example, the exact distances of the highest-SNR sources can vary among different DWD populations. Nevertheless, we expect the general trends described above to remain the same.

\section{Systematics\label{subsec:systematics}}

Here we discuss various effects not included in our analysis that may affect the binary evolution and the resulting gravitational waveform. Many sources of systematics relevant to our analysis are also found in the context of pulsar timing~\cite{Edwards:2006zg}.

\noindent
{\bf \textit{Tertiary perturber.}} While all DWDs in our Galaxy are under the influence of the overall Galactic gravitational potential, the local gravitational environment for a given DWD may be dominated by the presence of nearby tertiary objects. These could be any relatively small-scale inhomogeneities in the Galaxy, such as satellite galaxies, globular clusters, stars, or planets.

The possible effects of a single tertiary object on the GW signal emitted by a binary have been quantified in the past by several authors (see, e.g.,~\cite{Seto:2008di, Meiron:2016ipr,Tamanini:2018cqb,Robson:2018svj,Tamanini:2019usx,Xuan:2020xrr,Strokov:2021mkv,Seto:2023rfh}). In summary, the gravitational influence of a tertiary object causes the center of mass of the GW-emitting binary to follow a Keplerian orbit whose period can be shorter or longer than the 4-10 yr LISA observation time. Roughly speaking, in the former case the tertiary perturber will induce a distinct, periodic signal, which can in principle be modeled; while in the latter case the induced acceleration is nearly constant, and can mimic the Galactic acceleration if the magnitude of the induced acceleration is $\sim 10^{-10}\text{ m}/\text{s}^2$. A star of typical mass $m_3\sim M_\odot$ located at a typical distance $r_3\sim 1\text{ pc}$ from a DWD imparts only a tiny acceleration of order $m_3/r^2\sim 10^{-13}\text{ m}/\text{s}^2$. However, in a fraction of cases, stars or other objects lie close enough to yield significant accelerations to DWDs of interest during the LISA observation period. In particular, recent population-synthesis simulations suggest that $\mathcal{O}(1)$ fraction of nearby DWDs within 20 pc from the Sun are part of a hierarchical triple with a distant tertiary companion~\cite{2017A&A...602A..16T,Robson:2018svj}. That said, the statistics of three-body systems involving DWDs are yet to be confirmed--nor has the resulting effect on GW estimates of Galactic acceleration--since only a small number of such systems have been observed so far.

\noindent
{\bf \textit{Relativistic and propagation effects.}}
A key step in deriving the apparent GW phase, as discussed above, is mapping the proper time of the GW-emitting binary to the corresponding arrival time at the solar system. The pulsar timing community routinely takes into account several effects that can result in a dephasing of the associated EM signal, in particular: (1) \textit{Einstein delay}, i.e., the special relativistic and gravitational time dilation of the source's inner orbital period; (2) \textit{Romer delay}, i.e., the variation in the path length of light due to the source's motion relative to the solar system; and (3) \textit{Shapiro delay}, i.e., the delay in the arrival time of GWs when they are affected by the gravitational field of a massive body along their propagation path. 
The first time derivatives of these delays and the intrinsic frequency of the binary together determine the reference frequency $f_0$ of the GW signal, which, in the analysis above, we take to be the observed GW frequency at the beginning of the LISA observation. In other words, the first time derivative effects of these delays are absorbed into the definition of $f_0$.

\noindent
{\bf \textit{Non-gravitational interactions.}} We assume in our analysis that DWD binary evolution is driven purely by gravitational radiation. While most of the DWDs in the LISA band are expected to be compact and detached, some may undergo significant non-gravitational binary interactions~\cite{Littenberg:2018xxx}. Lower mass (i.e., less compact) WDs that reside in binaries with smaller orbital separations are prone to tidal deformations~\cite{Willems:2007nq,2014ApJ...791...76S} and mass transfers. Note, however, that although binary mass transfer contributes to systematic errors if not properly modeled, it can potentially be utilized to infer the individual masses of the WDs in a binary~\cite{Wolz:2020sqh, Yi:2023osk}, enabling inference of the chirp mass. The orbital evolution of DWDs may also be altered by EM radiation if at least one of the WDs is magnetic (about 20\% of the total white dwarf population is expected to have strong magnetic fields, in the range $10^6-10^9\text{ G}$~\cite{Bourgoin:2021yzf,Carvalho:2022pst,Marsh:2005tk,2018ApJ...868...19W}).
In addition, DWD interactions with ambient matter (and possibly dark matter) can contribute to the dissipation of their orbital energies~\cite{Chen:2020lpq,Caputo:2020irr,Pani:2015qhr,2017PhRvD..96f3001G}.  The aforementioned effects would complicate the orbital evolution beyond what is expected from a purely GW-driven binary orbit shrinkage. There are, in fact, examples of known DWDs in the current catalog of LISA verification binaries whose evolution is inadequately described by gravitational radiation alone~\cite{Marsh:2005tk,Munday:2022wtz}. A more complete modeling of the gravitational waveform should take into account such non-gravitational effects.

\section{Conclusions \label{sec:conclusion}}

In this paper we explore the prospect of using the $\mathcal{O}(10^{4})$ resolved double white dwarf (DWD) binaries that LISA is expected to observe as test mass accelerometers for probing the Milky Way gravitational potential. We assume a simple model for the Milky Way gravitational field profile composed of a Miyamoto-Nagai disk, a Hernquist bulge, and a Navarro-Frenk-White dark matter halo. In order to realistically assess whether this technique can work in practice, we construct an artifical catalog of LISA-detectable DWDs using the FZ binary population synthesis model of Ref.~\cite{Thiele:2021yyb}.
We let the DWD binary motion follow the Galactic rotation curve as defined by the balance between their centrifugal force and the Galactic potential gradient, neglecting their velocity dispersions. The apparent line-of-sight acceleration of each source, namely the sum of its actual line-of-sight acceleration (assumed to follow the Milky Way gravitational field model) and the perspective acceleration due to its proper motion (assumed to follow the Galactic rotation curve), cause time-dependent changes to the arrival time of the associated GW signals, which manifest as Doppler modulations in the measured GW phase.

We quantify LISA's sensitivity to the motions of these sources with a Fisher matrix analysis, both in the time-domain and in the frequency domain. We first derive the time-domain gravitational waveform by relating it to the source-frame phase via a mapping between the observation time and the source-frame time. We then obtain the frequency-domain waveform using the stationary phase approximation.
We perform the Fisher matrix analysis in both domains and find that the results agree. Our results show that strong correlations between the apparent acceleration of a source and its binary parameters undermine the ability to recover the apparent acceleration via GW-only measurements. This suggests the need of combining the GW analysis with other (independent) information about the binary parameters, e.g. by using detailed models of binary interactions~\cite{Wolz:2020sqh,Yi:2023osk} or EM observations of the same sources~\cite{Breivik:2017jip,Korol:2017qcx,Korol:2018wep,2014ApJ...791...76S}. If we assume that (1) EM observations can provide independent and accurate measurements of DWD chirp masses and (2) the EM counterparts to all these LISA sources are identified and observed, our analysis indicates that it is nearly possible to determine the normalization $\mathcal{N}$ of the Milky Way gravitational potential model. If all DWD parameters other than $\mathcal{N}$ are measured electromagnetically, then $\mathcal{N}$ can be determined after finding about $10^{3}$ EM counterparts. A more sophisticated analysis that accounts for correlations between the acceleration and position of a GW source in the Milky Way can further reduce measurement uncertainties. We also comment on possible sources of systematics that should be accounted for in more realistic future studies.

Previous works based on EM observations have utilized a variety of \textit{dynamic} tracers to study different aspects of the Milky Way. These include radial stellar acceleration measurements through precision spectrographs~\cite{Ravi:2018vqd, Silverwood:2018qra, Finan-Jenkin:2023hph, 2022ApJ...928L..17C}, angular stellar acceleration measurements through Gaia astrometry~\cite{Buschmann:2021izy}, peculiar acceleration measurements of globular clusters~\cite{Quercellini:2008it}, pulsar timing~\cite{NANOGrav:2015oil, Phillips:2020xmf,2021ApJ...907L..26C,2023arXiv230613137M,2021MNRAS.504..166H}, and the timing of eclipses in binaries~\cite{2022ApJ...928L..17C}. Different tracers have different kinematic distributions and biases, with some correlation to their age, metallicity, and location. These tracers offer different tradeoffs, and so provide complementary information when considered collectively.

\begin{acknowledgments}

We thank Andrea Antonelli for useful discussions. This work was supported by the Argonne National Laboratory under Award No. 2F60042; the Army Research Laboratory MAQP program under Contract No. W911NF-19–2-0181; and the University of Maryland Quantum Technology Center. E.B. and V.S. are supported by NSF Grants No.~AST-2006538, PHY-2207502, PHY-090003 and PHY-20043, by NASA Grants No.~20-LPS20-0011 and~21-ATP21-0010, by the Simons Foundation, and by the John Templeton Foundation Grant~62840. This work was also made possible through the support of Grant 63034 from the John Templeton Foundation. The opinions expressed in this publication are those of the authors and do not necessarily reflect the views of the John Templeton Foundation. E.B. and V.S. acknowledge support from the ITA-USA Science and Technology Cooperation program supported by the Ministry of Foreign Affairs of Italy (MAECI) and from the Indo-US Science and Technology Forum through the Indo-US Centre for Gravitational-Physics and Astronomy, grant~IUSSTF/JC-142/2019. E.H.T. acknowledges support by NSF Grant PHY-2310429 and Gordon and Betty Moore Foundation Grant No. GBMF7946. The authors also acknowledge the Texas Advanced Computing Center (TACC) at The University of Texas at Austin for providing HPC resources that have contributed to the research results reported within this paper~\cite{10.1145/3311790.3396656} (URL: \url{http://www.tacc.utexas.edu}). The authors further acknowledge the University of Maryland supercomputing resources (URL: \url{http://hpcc.umd.edu}) made available for conducting the research reported in this paper.

This research made use of the following software: IPython~\cite{2007CSE.....9c..21P}, SciPy~\cite{2020NatMe..17..261V},  Matplotlib~\cite{2007CSE.....9...90H}, NumPy~\cite{2011CSE....13b..22V}, SymPy~\cite{Meurer:2017yhf}, Astropy~\cite{2022ApJ...935..167A},
LEGWORK~\cite{2022ApJS..260...52W},
\texttt{mpmath}~\cite{mpmath}, \texttt{filltex}~\cite{2017JOSS....2..222G}.
\end{acknowledgments}

\appendix

\section{Perspective acceleration\label{app:shklovskii}}

For a small source displacement $\Delta D=|\vec{D}(\tau)-\vec{D}_0|\ll D_0$, we can approximately write
\beqa
    D&=&\sqrt{\left(D_0+\Delta D_\parallel\right)^2+\Delta D_\perp^2}\nonumber\\
    &\approx& D_0\left(1+\frac{\Delta D_\parallel}{D_0}+\frac{1}{2}\frac{\Delta D_\perp^2}{D_0^2}\right)\,, 
\eeqa
where $\Delta D_\parallel$ and $\Delta D_{\perp}$ are the source displacements parallel and perpendicular to the line of sight given by the initial position $\vec{D}_0$, and we have also neglected higher-order terms $\sim(\Delta D/D_0)^{3}$. If the source is moving with a constant acceleration $\mathbf{a}$ in its local frame,
\begin{align}
    \Delta \vec{D}(\tau)=\vec{v}_0 \tau+\frac{1}{2}\vec{a} \tau^2\,,
\end{align}
where $\mathbf{v}_0$ is the initial velocity in its local frame, then the change in the distance reads
\be
D(\tau) - D_0 = v_{\parallel,0}\tau + \frac 12\left(a_{\parallel,0} + \frac{v_{\perp,0}^2}{D_0}\right)\tau^2 + \mathcal{O}(\tau^3)\,, 
\ee
where the higher-order terms in time can be neglected, because a typical increment in the velocity during the observation time is much smaller than the Galactic velocity dispersion:
\beqa
\Delta v&\sim& aT_{\rm obs}\sim 10^{-10}\;\mbox{m}/\mbox{s}^2\times 10\;\mbox{yr}\nonumber\\
&\sim& 1\;\mbox{cm}/\mbox{s}\ll v_0\sim 100\;\mbox{km}/\mbox{s}\,. \label{eq:velocity_increment}
\eeqa

\section{Details of the Fisher matrix calculation\label{app:fisher}}

\subsection{Frequency-domain waveforms\label{app:subsec:FDwaveforms}}

Here we collect the FD counterparts of the sky-averaged TD waveforms used in this work. Recall that we use a capital~$F$ for the GW Fourier frequency, while $f=f(t)$ is reserved for the varying GW frequency in the TD. Also, the subscript~$F$ indicates a FD waveform which, in the stationary phase approximation, is obtained as follows:
\begin{widetext}
\beqa
h_F &=& \int\limits_0^{T_{\rm obs}}{\dd{t}\,A\left[f(t)\right]}\cos{\left(\psi_0+2\pi\int_0^t{\dd{t'}f(t')}\right)}e^{-2\pi i Ft} \nonumber \\
&=& \frac 12\int\limits_0^{T_{\rm obs}}{\dd{t}\,A\left[f(t)\right]}\exp\left\{i\psi_0+2\pi i\int_0^t{\dd{t'}f(t')}-2\pi i Ft\right\} + \frac 12\int\limits_0^{T_{\rm obs}}{\dd{t}\,A\left[f(t)\right]}\exp\left\{-i\psi_0-2\pi i\int_0^t{\dd{t'}f(t')}-2\pi i Ft\right\} \nonumber \\
&\approx & \frac 12 A\left[f(t_0)\right]\exp\left\{i\psi_0+2\pi i\int_0^{t_0}{\dd{t'}f(t')}-2\pi i Ft_0\right\}\int\limits_0^{T_{\rm obs}}{\dd{t}\,e^{i\pi \dot{f}(t_0)(t-t_0)^2}}\,, \label{app:eq:Fourier}
\eeqa
\end{widetext}
where $t_0$ is the stationary point,
\be 
f(t_0)=F\quad \Leftrightarrow \quad t_0 = t(F)\,.
\ee
The stationary phase approximation also implies that, as long as \mbox{$0\leq t_0\leq T_{\rm obs}$}, we can extend the limits of integration in the last integral to infinity, whereas outside of that range, the integral, and therefore, the FD waveform, vanishes. Computing the resulting Gaussian integral and integrating by parts in the phase, we obtain:
\be
h_F = \left\{
\begin{array}{ll}
\displaystyle\frac 12 \frac{A(F)}{\sqrt{\dot{f}\left[t(F)\right]}}e^{i\frac{\pi}{4}+i(\psi_0+\psi_F)}\,, & f_0\leq F\leq f(T_{\rm obs})\,, \\
{} & {} \\
0\,, & \mbox{otherwise}\,,
\end{array}
\right.
\ee
\be
\psi_F = -2\pi \int_{f_0}^{F}{\dd{f'}\,t(f')}\,, \quad f_0\equiv f(t=0)\,,
\ee
where $t(f)$ is the inverse of $f(t)$. For effects of windowing in the case of slowly chirping GW sources such as DWDs, see, for example, Appendix~D of Ref.~\cite{2023arXiv231200121S}.

In a phenomenological framework, the GW frequency is represented as a Taylor expansion in time. For the purposes of this paper, we are only interested in the following two cases:
\begin{itemize}
\item \underline{$\dot{f}={\rm const}$}:
\beqa
f = f_0 + \dot{f}_0 t  \quad&\Rightarrow&\quad t(f) = \frac{f-f_0}{\dot{f}_0}\,, \\
\psi_F &=& -\frac{\pi\left(F-f_0\right)^2}{\dot{f}_0}\,.
\label{app:eq:FDphase1}
\eeqa
\item \underline{$\ddot{f}={\rm const}$}:
\beqa
f &=& f_0 + \dot{f}_0 t + \frac 12\ddot{f}_0 t^2\,, \\
t(f) &=& \frac{-\dot{f}_0+\sqrt{\dot{f}_0^2+2\ddot{f}_0(f-f_0)}}{\ddot{f}_0} \nonumber \\
&\approx & \frac{f-f_0}{\dot{f}_0} - \frac 12\frac{\ddot{f}_0(f-f_0)^2}{\dot{f}_0^3}\,. \\
\psi_F &\approx& -\frac{\pi\left(F-f_0\right)^2}{\dot{f}_0} + \frac{\pi}{3}\frac{\ddot{f}_0\left(F-f_0\right)^3}{\dot{f}_0^3}\,.
\label{app:eq:FDphase2}
\eeqa
\end{itemize}

\subsection{Implementation\label{subsec:app:implement}}

In this section we provide details of our implementation of the Fisher matrix analysis. We also run a consistency check by comparing results from three versions of the calculation: the TD version, the FD version, and a version emulating the discrete nature of the LISA data stream.

To implement the calculation, we leverage the auto-differentiation and vectorization capabilities of the {\sc Python} package {\sc JAX}~\cite{jax2018github}. Auto-differentiation makes use of the well-known derivative rules to evaluate arbitrarily complicated derivatives on the fly to within machine precision. {\sc JAX} is also optimized for both CPUs and GPUs which, among other things, allows the user to vectorize a calculation, i.e., run it on all entries of an input array practically in parallel. 

These properties make {\sc JAX} a convenient tool for the Fisher matrix analysis on a population of DWDs. Indeed, the Fisher matrix for a GW signal, Eq.~(\ref{eq:FisherMatrix}), can be recast as the Hessian of the scalar inner product, Eq.~(\ref{eq:InnerProduct}), and the operation of computing and inverting the matrix can be vectorized and mapped across all DWDs in the population:
\beqa
F_{ij} &=& \frac 12\left.\frac{\partial^2}{\partial\theta_i\partial\theta_j}(\Delta h,\Delta h)\right|_{\boldsymbol{\theta}=\overline{\boldsymbol{\theta}}}\,, \label{app:eq:hessian}\\
\Delta h &\equiv& h(\boldsymbol{\theta})-h(\overline{\boldsymbol{\theta}})\,,
\eeqa
where $h(\boldsymbol{\theta})$ is the waveform as a function of the parameters~$\boldsymbol{\theta}=\{\theta_i\}$, and the derivatives are evaluated at the fiducial values $\overline{\boldsymbol{\theta}}$ of the parameters. As a side note, this Hessian form of the Fisher matrix is related to the linear-signal or high-SNR approximations (see e.g.~\cite{Vallisneri:2007ev}). 

In a more realistic version, the integral is substituted for a sum over the data points, which takes into account the fact that the LISA data have to be discretely sampled with a timestep $\Delta t$. We assume $\Delta t=1$~s. This results in a dataset of $T_{\rm obs}/\Delta t$ data points for a single datastream ($\approx 3\times 10^8$ data points for the cadence~$\Delta t=1$~s and $T_{\rm obs}=10$~yr).

We now provide the explicit expressions used in the TD version, the FD version, and the discretized version.

\noindent
{\bf \textit{Time domain.}} The TD waveform~$h(t)$, Eqs.~(\ref{eq:WaveformTD}) and~(\ref{eq:PhaseTD}), completes $\sim 10^5$ oscillations during the observation time, and the integration of such an oscillatory function can be challenging. For this reason we simplify the TD version of the Fisher matrix before we vectorize and evaluate it with {\sc JAX}. The TD Fisher matrix reads:
\begin{align}
        F_{ij}& = \mathcal{C}^2\int\limits_0^{T_{\rm obs}}{\frac{\partial_{i}h(t)\partial_{j}h(t)}{S_n[f(t)]}\,\dd{t}}\nonumber\\
        &\approx \frac{\mbox{SNR}^2}{A^2T_{\rm obs}}\int\limits_{0}^{T_{\rm obs}}{2(\partial_{i}h)(\partial_{j}h)\,\dd{t}}\,,
\end{align}
where partial derivatives are taken w.r.t. the components of the parameter vector $\boldsymbol{\theta}=\{\ln A,\ln f_0,\ln\dot{f}_0,\ln\ddot{f}_0,\psi_0\}$. If we neglect terms $\mathcal{O}\left(1/f_0T_{\rm obs}\right)$, the matrix is effectively reduced to a $4\times 4$ block. Indeed, for the cross-terms with $\ln A$ ($\theta_i\neq\theta_j=\ln A$),
\be
F_{ij}\approx \mbox{SNR}^2\times \mathcal{O}\left(\frac{\partial_{\theta_i}\psi}{f_0T_{\rm obs}}\right)\approx 0\,, \label{eq:app:FisherTDoff-diag}
\ee
and the full matrix
\be
F_{ij} \approx \mbox{SNR}^2\times\left(
\begin{array}{cc}
1 & \boldsymbol{0} \\
{} & {} \\
\boldsymbol{0} & F_{IJ}
\end{array}
\right)\,, \label{eq:app:FisherTDcontinuous}
\ee
where $\theta_{I,J}\neq\ln A$, and the $3\times 3$ block
\beqa
F_{IJ}&\equiv&\int_0^{T_{\rm obs}}{(\partial_{I}\psi)(\partial_{J}\psi)\frac{\dd{t}}{T_{\rm obs}}} \nonumber \\
&=& \frac{1}{2T_{\rm obs}}\left[\frac{\partial^2}{\partial\theta_i\partial\theta_j}\int\limits_0^{T_{\rm obs}}{(\Delta\psi)^2\dd{t}}\right]_{\boldsymbol{\theta}=\overline{\boldsymbol{\theta}}}\,,
\eeqa
with $\Delta\psi\equiv \psi(\boldsymbol{\theta})-\psi(\overline{\boldsymbol{\theta}})$\,.

\noindent
{\bf \textit{Frequency domain.}} The inner product that goes into Eq.~(\ref{app:eq:hessian}) reads
\beqa
(\Delta h_F,\Delta h_F) &=& \frac 12\mbox{SNR}^2 \nonumber \\
&\times& \int\limits_{-1}^{1}{\left|e^{\ln A - (\ln\dot{f}_0 - \ln\overline{\dot{f}}_0)/2}e^{i\psi_F(\boldsymbol{\theta})}-e^{i\psi_F(\overline{\boldsymbol{\theta}})}\right|^2\dd{x}}\,, \nonumber \\
&{}&
\eeqa
\beqa
F &=& f_0 + \frac 12(x+1)\Delta f\,, \quad \Delta f\equiv f(T_{\rm obs}) - f_0\,,
\eeqa
where the phase is given by Eq.~(\ref{app:eq:FDphase2}), and we have recast the integral in a form convenient for Gauss--Legendre quadrature. Note that we also neglect a variation in $\dot{f}$ in the amplitude, which would result in a correction of the order of $T_{\rm obs}/\tau_{\rm c}\sim 10^{-5}$.

\textbf{Discretized time domain}. A discrete version of the TD inner product reads:
\beqa
(\Delta h,\Delta h) &\approx& \mbox{SNR}^2\sum\limits_{p=0}^{N}\left(e^{\ln A}\cos{\psi_p(\boldsymbol{\theta})}-\cos{\psi_p(\overline{\boldsymbol{\theta}})}\right)^2\,, \nonumber \\
& {} &
\eeqa
where $\psi_p$ is a TD phase evaluated at moments of time \mbox{$t_p=p\Delta t$}, and the number of datapoints $N$ is given by the integer part of $T_{\rm obs}/\Delta t$. Again, we assume that the noise $S_{\rm n}[f(t_p)]\approx{\rm const}$ for the quasimonochromatic sources considered in this paper. In the discrete TD version, we no longer neglect the off-diagonal terms as we did in the continuous case above, see Eqs.~(\ref{eq:app:FisherTDoff-diag}) and~(\ref{eq:app:FisherTDcontinuous}).

We compare Fisher matrix uncertainties obtained in the three approaches described above, and evaluated on the synthetic population of DWDs adopted in this paper. We find the following relative discrepancies (that can be thought of as ``uncertainty on the uncertainties''):
\begin{itemize}
\item \underline{$\dot{f}={\rm const}$}, TD vs. FD: $\approx 10^{-11.5}$ across the parameter set $\{\ln A,\ln{f_0},\ln{\dot{f}_0}, \psi_0\}$.
\item \underline{$\dot{f}={\rm const}$}, TD vs. discretized TD: $\approx 10^{-7}$ for the same parameters.  
\item \underline{$\ddot{f}={\rm const}$}, TD vs. FD: $\approx 10^{-5}$ for the variables $\{\ln{f_0},\ln{\dot{f}_0}, \ln{\ddot{f}_0}, \psi_0\}$, and $\approx 10^{-11}$ for $\ln A$. 
\end{itemize}
In each of these cases, the scatter is $\approx 0.5$~dex (i.e., less than an order of magnitude).

We also rely on JAX to switch between the sets of parameters $\boldsymbol{\theta}^{(1)}$, $\boldsymbol{\theta}^{(2)}$, $\boldsymbol{\theta}^{(3)}$ listed in Section~\ref{subsec:FD}. As is known, the results of a Fisher matrix calculation in one set of parameters~$\theta_i$ can be converted to another set~$\widetilde{\theta}_j$ by applying the Jacobian matrix $\widetilde{J}_{ij}\equiv\partial\widetilde{\theta}_i/\partial\theta_j$. Being a matrix of derivatives, the Jacobian naturally fits into the JAX auto-differentiation logic, as do the matrix transforms for a nonsingular Fisher matrix~$F$ and the corresponding covariance matrix~$\Sigma=F^{-1}$:
\be
\widetilde{\Sigma} = \widetilde{J}\Sigma\widetilde{J}^{\rm T}\,, \qquad \widetilde{F} = \left(\widetilde{J}^{-1}\right)^{\rm T}F\widetilde{J}^{-1}\,.
\ee
All those operations are also easily vectorized in JAX.

\bibliography{references}

\end{document}